\documentclass[aps,prl,twocolumn,superscriptaddress,showpacs,preprintnumbers,amsmath,amssymb]{revtex4-1}

\usepackage{longtable}
\usepackage{amsmath}
\usepackage{graphicx} 
\usepackage{dcolumn}  

\begin{document}


\preprint{\vbox{ \hbox{   }
			\hbox{Belle Preprint {\it 2015-1}}
                        \hbox{KEK Preprint {\it 2014-42}}
                        \hbox{Accepted to {\it JHEP}}
}}

\title{ \quad\\[1.0cm] Search for $B$ decays to final states with the $\eta_c$ meson}

\author{A.~Vinokurova}\affiliation{Budker Institute of Nuclear Physics SB RAS and Novosibirsk State University, Novosibirsk 630090, Russian Federation}
\author{A.~Kuzmin}\affiliation{Budker Institute of Nuclear Physics SB RAS and Novosibirsk State University, Novosibirsk 630090, Russian Federation}
\author{S.~Eidelman}\affiliation{Budker Institute of Nuclear Physics SB RAS and Novosibirsk State University, Novosibirsk 630090, Russian Federation}
\author{A.~Abdesselam}\affiliation{Department of Physics, Faculty of Science, University of Tabuk, Tabuk 71451, Saudi Arabia}
\author{I.~Adachi}\affiliation{High Energy Accelerator Research Organization (KEK), Tsukuba 305-0801, Japan}\affiliation{The Graduate University for Advanced Studies, Hayama 240-0193, Japan}
\author{H.~Aihara}\affiliation{Department of Physics, University of Tokyo, Tokyo 113-0033, Japan}
\author{K.~Arinstein}\affiliation{Budker Institute of Nuclear Physics SB RAS and Novosibirsk State University, Novosibirsk 630090, Russian Federation}
\author{D.~M.~Asner}\affiliation{Pacific Northwest National Laboratory, Richland, WA 99352, USA}
\author{T.~Aushev}\affiliation{Moscow Institute of Physics and Technology, Moscow Region 141700, Russian Federation}\affiliation{Institute for Theoretical and Experimental Physics, Moscow 117218, Russian Federation}
\author{R.~Ayad}\affiliation{Department of Physics, Faculty of Science, University of Tabuk, Tabuk 71451, Saudi Arabia}
% ...
\collaboration{The Belle Collaboration}
\noaffiliation

\begin{abstract}
We report a search for $B$ decays to selected final states with the $\eta_c$ meson: 
$B^{\pm}\to K^{\pm}\eta_c\pi^+\pi^-$, $B^{\pm}\to K^{\pm}\eta_c\omega$, 
$B^{\pm}\to K^{\pm}\eta_c\eta$ and $B^{\pm}\to K^{\pm}\eta_c\pi^0$. The analysis 
is based on $772\times 10^6$ $B\bar{B}$ pairs collected at the 
$\Upsilon(4S)$ resonance with the Belle detector at the KEKB 
asymmetric-energy $e^+e^-$ collider.
We set 90\% confidence level upper limits on the branching fractions of the studied $B$ decay modes, 
independent of intermediate resonances, in the range $(0.6-5.3)\times 10^{-4}$. 
We also search for molecular-state candidates in the 
$D^0\bar{D}^{*0}-\bar{D}^0D^{*0}$, $D^0\bar{D}^0+\bar{D}^0D^0$ and 
$D^{*0}\bar{D}^{*0}+\bar{D}^{*0}D^{*0}$ combinations, neutral partners of the 
$Z(3900)^{\pm}$ and $Z(4020)^{\pm}$, and a poorly understood state $X(3915)$ 
as possible intermediate states in the decay chain, and set 90\% confidence level 
upper limits on the product of branching fractions to the mentioned intermediate states and decay 
branching fractions of these states in the range $(0.6-6.9)\times 10^{-5}$.
\end{abstract}

\maketitle
\tighten
{\renewcommand{\thefootnote}{\fnsymbol{footnote}}}
\setcounter{footnote}{0}

\section{Introduction}

Many exotic charmonium-like states are observed in the mass region above the $D\bar{D}$ threshold. Decays of $B$ mesons provide a fruitful opportunity to study these states and to find new ones.
For example, the state $X(3872)$ was first observed by Belle in exclusive $B^+\to K^+\pi^+\pi^-J/\psi$ decays~\cite{Belle1} that was later confirmed by  CDF~\cite{CDF1}, D\O~\cite{D01} and BaBar~\cite{BaBar1}. It was also observed in the LHCb experiment~\cite{LHCb1,LHCb2} in $pp$ collisions and $B$ decays. The $X(3872)$ mass is close to the $m_{D^0}+m_{\bar{D}^{*0}}$ threshold, which engendered a hypothesis that this state may be a $D^0\bar{D}^{*0}$ molecule~\cite{Molecule}. The observation of the decay $X(3872)\to\gamma J/\psi$ by BaBar~\cite{BaBar2} and Belle~\cite{Belle2} established the charged parity of $X(3872)$ to be positive. Angular analysis of the $X(3872)\to J/\psi\pi^+\pi^-$ decay by LHCb~\cite{LHCb2} determined all its quantum numbers: $J^{PC}=1^{++}$.

If $X(3872)$ is indeed a $D^0\bar{D}^{*0}$ molecule, there can exist other ``$X(3872)$-like'' molecular states with different quantum numbers. Some may reveal themselves in the decays to final states containing the $\eta_c$ meson. For example, a $D^0\bar{D}^{*0}-\bar{D}^0D^{*0}$ combination (denoted hereinafter by $X_1(3872)$) with quantum numbers $J^{PC}=1^{+-}$ would have a mass around $3.872$ GeV/$c^2$ and would decay to $\eta_c\rho$ and $\eta_c\omega$. Combinations of $D^0\bar{D}^0+\bar{D}^0D^0$, denoted by $X(3730)$, and $D^{*0}\bar{D}^{*0}+\bar{D}^{*0}D^{*0}$, denoted by $X(4014)$, with quantum numbers $J^{PC}=0^{++}$ would decay to $\eta_c\eta$ and $\eta_c\pi^0$. The mass of the $X(3730)$ state would be around $2m_{D^0}=3.730$ GeV/$c^2$ while that of the $X(4014)$ state would be near $2m_{D^{*0}}=4.014$ GeV/$c^2$.

Recently, a new charged state $Z(3900)^{\pm}$ was found in $Y(4260)$ decays by Belle~\cite{zbelle} and BESIII~\cite{zbes1}. Since this particle is observed in the decay to $\pi^{\pm}J/\psi$, it should contain at least four quarks. BESIII~\cite{zbes2} reported subsequently an observation of another decay channel of the seemingly same state $Z(3885)^{\pm}\to (D\bar{D}^*)^{\pm}$. The $Z(3900)^{\pm}$ was confirmed in the decay to $\pi^{\pm}J/\psi$ by an analysis of CLEO-c data~\cite{zcleo} that also reported evidence for its neutral isotopic partner $Z(3900)^0$.
Another exotic charged state $Z(4020)^{\pm}$ was observed by BESIII in decays to $\pi^{\pm}h_c$~\cite{zbes3} and $(D^*\bar{D}^*)^{\pm}$~\cite{zbes4}.
There are some indications from these analyses that the spin and parity of the charged states might be $J^P=1^+$.

The near-threshold enhancement in the $\omega J/\psi$ invariant mass distribution named $Y(3940)$ was first observed by Belle in exclusive $B\to K\omega J/\psi$ decays~\cite{x39151}. Later, in the same decay mode, BaBar discovered $X(3915)$~\cite{x39152}, which was confirmed by Belle in two-photon production~\cite{x39153} and by other BaBar measurements~\cite{x39154, x39155}. The parameters of $Y(3940)$ are consistent with those of $X(3915)$, so they are considered to be the same particle. The quantum numbers of $X(3915)$ are claimed to be $J^{PC}=0^{++}$, but its nature is still undetermined, and there are several interpretations describing this state~\cite{x39156,x39157,x39158,x39159,x391510}. 

To search for the particles described above, we reconstruct $\eta_c$ mesons via the $K_S^0K^{\pm}\pi^{\mp}$ mode and study the following four decays of charged $B$ mesons:
\begin{enumerate}
\item the ($\pi^+\pi^-$) decay mode: $B^{\pm}\to K^{\pm}X\to K^{\pm}(\eta_c\pi^+\pi^-)$,
where we look for $X_1(3872)$, $Z(3900)^0$ and $Z(4020)^0$;
\item the ($\omega$) decay mode: $B^{\pm}\to K^{\pm}X\to K^{\pm}(\eta_c\omega)$,
where we look for $X_1(3872)$;
\item the ($\eta$) decay mode: $B^{\pm}\to K^{\pm}X\to K^{\pm}(\eta_c\eta)$,
where we look for $X(3730)$, $X(4014)$ and $X(3915)$;
\item the ($\pi^0$) decay mode: $B^{\pm}\to K^{\pm}X\to K^{\pm}(\eta_c\pi^0)$,
where we look for $X(3730)$, $X(4014)$ and $X(3915)$.
\end{enumerate}

\section{Event selection}

The analysis is based on a data sample that contains $772\times 10^6$
$B\bar{B}$ pairs, collected with the Belle detector at the KEKB 
asymmetric-energy $e^+e^-$ collider~\cite{KEKB1,KEKB2} operating at the $\Upsilon(4S)$
resonance.

 The Belle detector~\cite{Belle11} (also see detector section in~\cite{Belle22}) is a large-solid-angle magnetic spectrometer
that consists
of a silicon vertex detector, 
a 50-layer central drift chamber
(CDC) for charged particle tracking and specific ionization measurement 
($dE/dx$), an array of aerogel threshold Cherenkov counters (ACC), 
time-of-flight scintillation counters (TOF), and an array of 8736 CsI(Tl) 
crystals for electromagnetic calorimetry located inside a superconducting
solenoid coil that provides a 1.5~T magnetic field. An iron flux return yoke located
outside the coil is instrumented to detect $K^0_L$ mesons and identify muons. 
We use a GEANT3-based Monte Carlo (MC) simulation to
model the response of the detector and determine its acceptance~\cite{sim}.

Charged tracks are selected with requirements based on the goodness of fit of the
tracks and their impact parameters relative to the interaction point (IP). We
require that the polar angle of each track be in the angular range
$(18^{\circ}-152^{\circ})$
and that the track momentum perpendicular to the positron beamline be greater 
than 100 MeV/$c$.

Pions and kaons are distinguished by combining the responses of 
the ACC and the TOF with $dE/dx$ measurements in the CDC
 to form a likelihood ratio $\mathcal{L}_{i/j}=\mathcal{L}_i/(\mathcal{L}_i+\mathcal{L}_j)$. Here, $\mathcal{L}_i$ is the likelihood that the particle is of type $i$.
$K^0_S$ candidates are reconstructed via the $\pi^+\pi^-$ final state. The $\pi^+\pi^-$ invariant mass must lie in the range $0.486$ GeV/$c^2$ $<M(\pi^+\pi^-)<0.510$ 
GeV/$c^2$. 
The flight length of the $K^0_S$ is required to lie within the interval 
$(0.1-20)$ cm. The angle $\varphi$ between the pion pair 
momentum and the line joining the $\pi^+\pi^-$ the vertex to the IP must satisfy $\cos\varphi>0.95$.

The invariant masses of intermediate resonances must lie within ranges obtained from signal MC and containing more than 95\% of the signal yield: $2.9254$ GeV/$c^2$ $<M(K^0_SK^{\pm}\pi^{\mp})<3.0454$ GeV/$c^2$ for the $\eta_c$ meson, $0.758$ GeV/$c^2$ $<M(\pi^+\pi^-\pi^0)<0.808$ GeV/$c^2$ for the $\omega$ meson, $0.125$ GeV/$c^2$ $<M(\gamma\gamma)<0.145$ GeV/$c^2$ for the $\pi^0$ meson, $0.528$ GeV/$c^2$ $<M(\gamma\gamma)<0.568$ GeV/$c^2$ and $0.538$ GeV/$c^2$ $<M(\pi^+\pi^-\pi^0)<0.558$ GeV/$c^2$ for the $\eta$ meson. 

$B$ candidates are identified 
by their center-of-mass (c.m.)\ energy difference 
$\Delta E=(\sum_iE_i)-E_{\rm b}$, and the
beam-constrained mass 
$M_{\rm bc}=\sqrt{E^2_{\rm b}/c^2-(\sum_i\vec{p}_i)^2}/c$, where 
$E_{\rm b}=\sqrt{s}/2$ is the beam energy in the $\Upsilon(4S)$
c.m.\ frame, and $\vec{p}_i$ and $E_i$ are the c.m.\ three-momenta and 
energies, respectively, of the $B$ candidate decay products. We define the signal region by $|\Delta E|<0.02$ GeV and $5.273$ GeV/$c^2$ $<M_{\rm bc}<5.285$ GeV/$c^2$; the $|\Delta E|$ range extends to $0.04$ GeV for some decay channels. 

Although the continuum background ($e^+e^-\to q\bar{q}$,
where $q=u,d,s,c$) is not dominant, we suppress it using topological criteria. Since  
the produced $B$ mesons 
are nearly at rest in the c.m.\ frame, the signal
tends to be 
isotropic, while the continuum $q\bar{q}$ background tends
to have  a two-jet structure. We use the angle $\Theta_{\rm thrust}$ between the thrust axis\footnote{I.e., axis $\hat{n}$ that maximizes $\Sigma_i|\hat{p_i}\cdot\hat{n}|$, where the sum is over all considered particles momenta $\hat{p_i}$.} of 
the $B$ candidate and that of the rest of the event
to discriminate between these two cases. The distribution of
$|\cos\Theta_{\rm thrust}|$ is strongly peaked near $1$
for $q\bar{q}$ events but nearly uniform for  $\Upsilon(4S)\to
B\bar{B}$ events; we require $|\cos\Theta_{\rm thrust}|<0.8$. In addition, we require that the c.m. polar angle $\theta_B$ of the reconstructed $B$ meson satisfies $|\cos\theta_B|<0.8$, where this angle is measured relative to the $z$ axis that is collinear with the positron beam.

The mean number of multiple $B$ candidates per event varies from $1.2$ to $2.0$, depending on the decay channel.
If an event has multiple $B$ candidates, we select the candidate with the minimum value of the following expressions in the order described below:
\begin{enumerate}
\item $|m_{K^0_S}-M(\pi^+\pi^-)|$, 
\item $|m_{\eta_c}-M(K^0_SK^{\pm}\pi^{\mp})|$, 
\item $|m_{\eta}-M(\gamma\gamma)|$ (for $\eta\to\gamma\gamma$ in the ($\eta$) mode) or $|m_{\pi^0}-M(\gamma\gamma)|$ (for the ($\omega$) mode, for $\eta\to\pi^+\pi^-\pi^0$ in  the ($\eta$) mode and for the ($\pi^0$) mode),
\item maximum difference between $z$ coordinates at the closest-distance point to the $z$ axis among each pair of charged particles in the signal-$B$ final state. 
\end{enumerate} 
According to MC, this procedure selects the correct $B$ candidate with 95\% probability.

\section{Reconstruction of final states}

\subsection{$B^{\pm}\to K^{\pm}\eta_c+{\rm hadrons}$}

\begin{figure}[htb]
\centering
\includegraphics[height=4 cm]{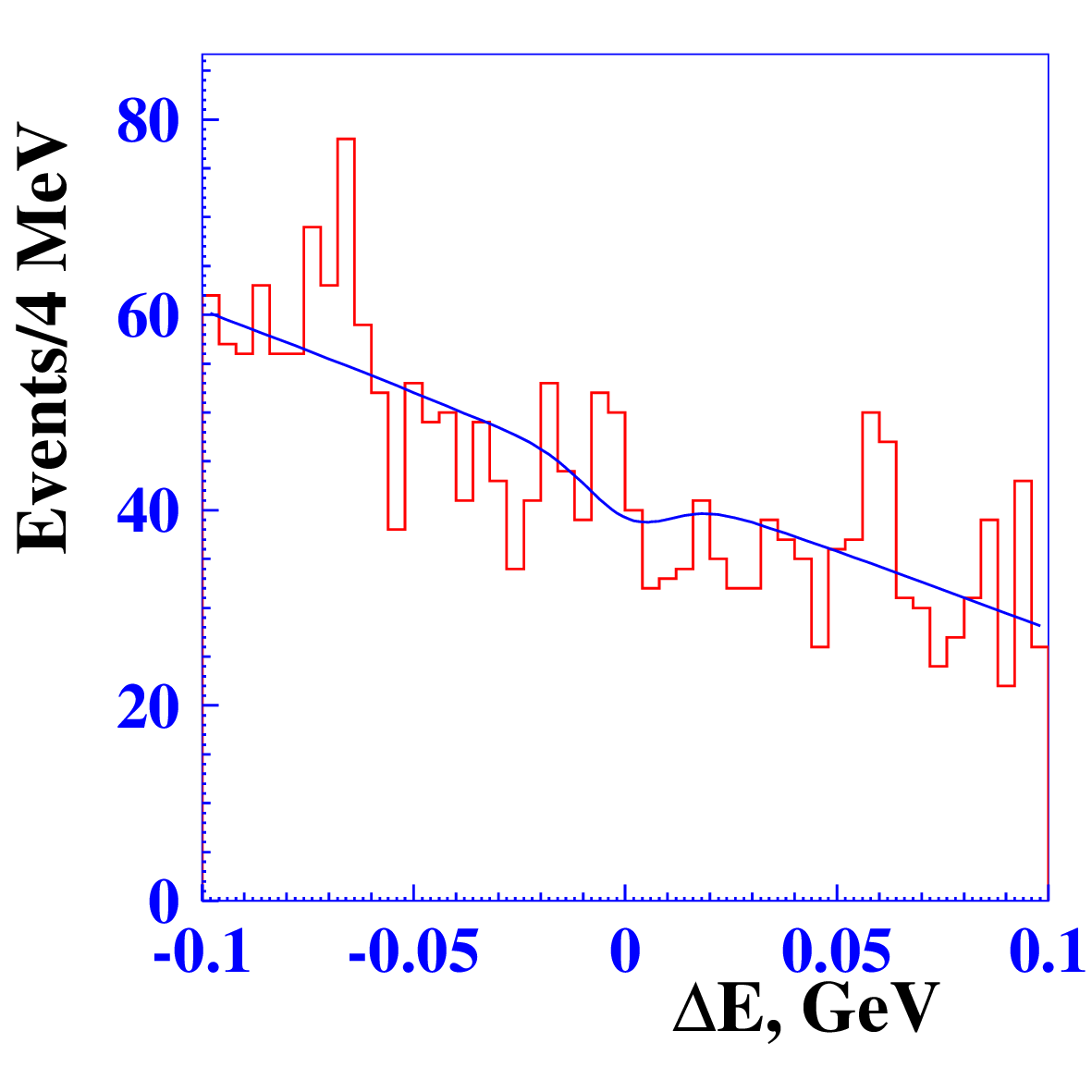}
\caption{The $\Delta E$ distribution for the decay $B^{\pm}\to K^{\pm}\eta_c\omega$.}
\label{pic:23}
\end{figure}

We search for ${D}^{(*)0}D^{(*)0}$ molecular-state candidates $Z(3900)^0$, $Z(4020)^0$, and $X(3915)$ in the following $B$ meson decays: $B^{\pm}\to K^{\pm}\eta_c\pi^+\pi^-$, $B^{\pm}\to K^{\pm}\eta_c\omega$, $B^{\pm}\to K^{\pm}\eta_c\eta$, and $B^{\pm}\to K^{\pm}\eta_c\pi^0$. 
To determine the branching fractions, we perform a binned maximum-likelihood fit of the $\Delta E$ distribution that is modelled by a peaking signal and featureless background.
For the ($\pi^+\pi^-$), ($\omega$) and ($\pi^0$) decay modes, the signal function is the sum of two Gaussians ($G$) 
and the background function is a linear polynomial; the fitting function is
\begin{equation}
f(x)=N_s\left[\alpha G(x_{1},\sigma_1)+(1-\alpha)G(x_{2},\sigma_2)\right]+c_0+c_1x.
\label{eq:21}
\end{equation}
For a given mode, we generate signal MC and fix the mean values $x_1$ and $x_2$, the standard deviations $\sigma_1$ and $\sigma_2$, and the  fraction of the first Gaussian $\alpha$; we also obtain the detection efficiency for the mode.
Here and in the following the detector resolution is taken into account.  
To account for the difference in resolution between 
MC and data, we replace the resolution of the dominant second Gaussian in the fit with 
$\sigma_2^{\prime} =\sqrt{\sigma_2^2+\delta^2}$, where the resolution degradation $\delta=(7.1\pm 2.3)$ MeV/c$^2$ is taken from the analysis of the decay $B^{\pm}\to K^{\pm}\eta_c(2S)\to K^{\pm}(K^0_SK^{\pm}\pi^{\mp})$~\cite{Vinok}. The $\Delta E$ distribution for the ($\omega$) mode is shown in figure~\ref{pic:23}.

\begin{figure}[htb]
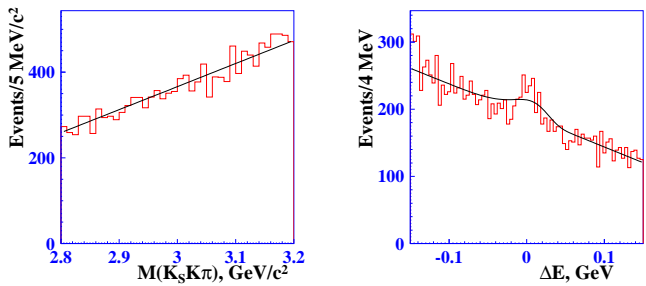

\centering
\includegraphics[height=4 cm]{fig3a.eps}
\hfill
\includegraphics[height=4 cm,origin=c,angle=0]{fig3b.eps}
\caption{Projections of the two-dimensional fit in $K_SK\pi$ invariant mass (left) and $\Delta E$ (right) for the decay $B^{\pm}\to K^{\pm}\eta_c\pi^0$. Each projection is plotted for events in the whole fitting range of the other projection.}
\label{pic:25}
\end{figure}

In the ($\pi^+\pi^-$) and ($\pi^0$) modes, we observe some significant signal and so perform a two-dimensional fit of the $K_SK\pi$ invariant mass ($x$) and $\Delta E$ ($y$) distributions:
\begin{equation}
f(x,y) = N_su(x)v(y) + N_\textnormal{\scriptsize non-res}v(y) +c_0+c_1x+c_2y, 
\label{eq:22}
\end{equation}
\begin{equation}
u(x) = b(M_{\eta_c},\Gamma_{\eta_c})\otimes G(0,\sigma_{\rm res}),
\label{eq:23a}
\end{equation}
\begin{equation}
v(y)=\left\{
\begin{array}{l}
\alpha G(y_1,\sigma_1)+(1-\alpha)G(y_2,\sigma_2)\\
\\
G_{\rm LG}(y_0,\sigma,P)\\
\end{array} \right.
\label{eq:23b24}
\end{equation}
for the ($\pi^+\pi^-$) and ($\pi^0$) modes, respectively.
The logarithmic Gaussian function is defined as 
\begin{eqnarray}
G_{\rm LG}(x_0,\sigma,P)=\frac{P}{\sqrt{2\pi}\sigma\sigma_0}e^{-\frac{{\rm ln}^2(1-P(x-x_0)/\sigma)}{2\sigma_0^2}-\frac{\sigma_0^2}{2}}, \nonumber 
\label{eq:100}
\end{eqnarray}
where $\sigma_0=\frac{1}{\sqrt{2{\rm ln}2}}\sinh^{-1}(P\sqrt{2{\rm ln}2})$ and $P$ is the asymmetry parameter.
The function $u(x)$ characterizes the $\eta_c$ resonance and is described by the convolution of a Breit-Wigner ($b$) function and a Gaussian detector resolution function with $\sigma_{\rm res}=6.2$ MeV/$c^2$ obtained from Ref.~\cite{Vinok}. 
The parameter $N_\textnormal{\scriptsize non-res}$ represents the number of events that do not contain an intermediate $\eta_c$ meson, but have the same final state. According to the fit, most of the events in the $\Delta E$ peak are of that origin. The $M(K_SK\pi)$ and $\Delta E$ distributions for ($\pi^+\pi^-$) and ($\pi^0$) modes are shown in figures~\ref{pic:22} and \ref{pic:25}, respectively.
\begin{figure}[htb]
\centering
\includegraphics[height=4 cm]{fig2a.eps}
\hfill
\includegraphics[height=4 cm,origin=c,angle=0]{fig2b.eps}
\caption{Projections of the two-dimensional fit in $K_SK\pi$ invariant mass (left) and $\Delta E$ (right) for the decay $B^{\pm}\to K^{\pm}\eta_c\pi^+\pi^-$. Each projection is plotted for events in the whole fitting range of the other projection.}
\label{pic:22}
\end{figure}

\begin{figure}[htb]
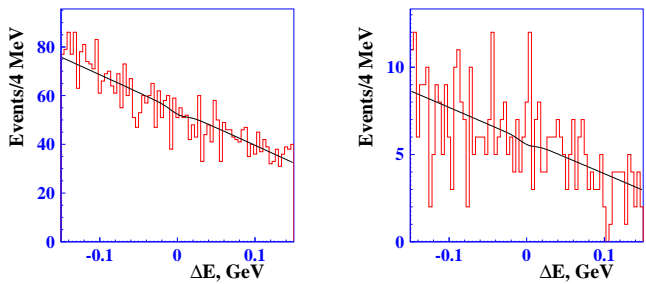

\centering
\includegraphics[height=4 cm]{fig4a.eps}
\hfill
\includegraphics[height=4 cm,origin=c,angle=0]{fig4b.eps}
\caption{The combined fit projections of the $\Delta E$ distributions in case of the $\eta\to\gamma\gamma$ (left) and $\eta\to\pi^+\pi^-\pi^0$ (right) modes for the decay $B^{\pm}\to K^{\pm}\eta_c\eta$.}
\label{pic:24}
\end{figure}

To improve the $\Delta E$ resolution in the ($\eta$) mode, we modify the energy of the $\eta$ candidate decaying into photons: $E_{\eta}^{\prime}=c\sqrt{m_{\eta}^2c^2+p_{\eta}^2}$, where $m_{\eta}=547.853$ MeV/c$^2$~\cite{PDG} is the mass and $p_{\eta}$ is the reconstructed momentum. Since the $\eta$ candidate is reconstructed in two decay modes, we perform a combined fit of the $\Delta E$ distribution corresponding to $\eta\to\gamma\gamma$ and $\eta\to\pi^+\pi^-\pi^0$, using the following function:
\begin{eqnarray}
f_i(x)&=&N_{\rm eff}\varepsilon_{i}{\mathcal B}_{i}\left[\alpha_iG(x_{1,i},\sigma_{1,i})+(1-\alpha_i)G(x_{2,i},\sigma_{2,i})\right] + \nonumber\\
& & c_{0,i}+c_{1,i}x,
\label{eq:25}
\end{eqnarray} 
where $i$ refers to either $\eta\to\gamma\gamma$ or $\eta\to\pi^+\pi^-\pi^0$ decay. In particular, ${\mathcal B}_{2\gamma}={\mathcal B}(\eta\to\gamma\gamma)$ and ${\mathcal B}_{3\pi}={\mathcal B}(\eta\to\pi^+\pi^-\pi^0)\times{\mathcal B}(\pi^0\to\gamma\gamma)$. Here, the parameter $N_{\rm eff}$ is the effective number of signal events. To obtain the total yield for each decay channel, it should be multiplied by the corresponding efficiency $\varepsilon_{2\gamma/3\pi}$ and $\eta$ decay branching fraction ${\mathcal B}_i$. The combined fit projections of the $\Delta E$ distributions for the ($\eta$) mode are shown in figure~\ref{pic:24}.

The fit results are summarized in table~\ref{tab:21}.
\begin{table}[htb]
\centering
\begin{tabular}{|c||c|c|c|} 
\hline
Decay mode & Fit. function & Efficiency, \% & Yield \\
\hline \hline
$B^{\pm}\to K^{\pm}\eta_c\omega$ & (\ref{eq:21}) & $0.53\pm 0.01$ & $-41\pm 27$\\
\hline
$B^{\pm}\to K^{\pm}\eta_c\pi^+\pi^-$ & (\ref{eq:22})+(\ref{eq:23a})+(\ref{eq:23b24}) & $2.84\pm 0.02$ &
$155\pm 72$ \\
\cline{1-1}
\cline{3-4}
$B^{\pm}\to K^{\pm}\eta_c\pi^0$ & & $3.69\pm 0.01$ & $-1.9\pm 12.1$ \\
\hline
$B^{\pm}\to K^{\pm}\eta_c\eta$, & & & \\
\cline{3-4}
$\eta\to\gamma\gamma$ & (\ref{eq:25}) & $3.05\pm 0.01$ & $-14\pm 26$ \\
\cline{3-4}
$\eta\to\pi^+\pi^-\pi^0$ &  & $ 0.69\pm 0.01$ & $-1.8\pm 3.4$ \\
\hline
\end{tabular}
\caption{\label{tab:21} Fit results for $B$ decays independent of intermediate resonances.}
\end{table}

\subsection{$X_1(3872)$, $X(3730)$ and $X(4014)$}

\begin{figure}[htb]
\centering
\includegraphics[height=4 cm]{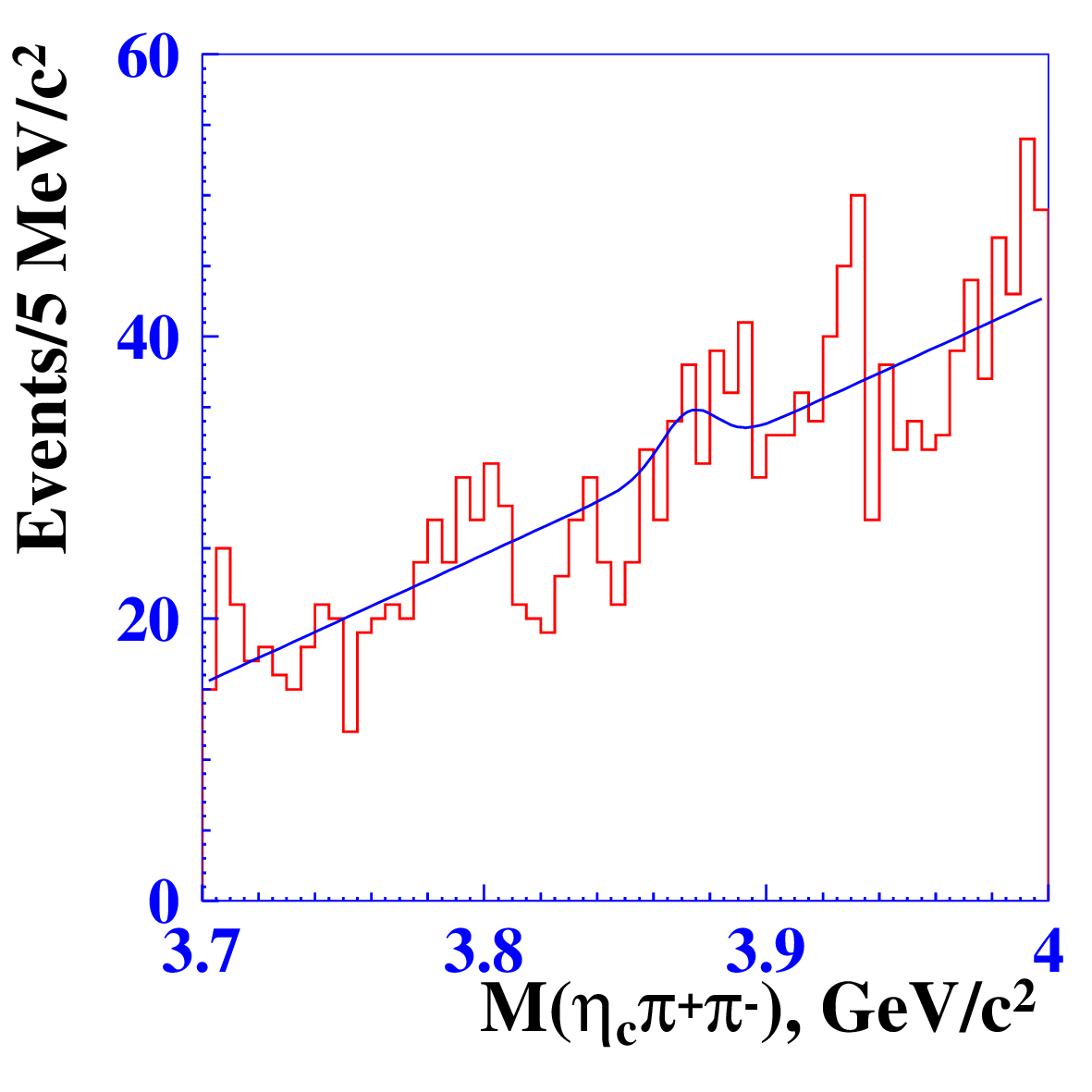}
\hfill
\includegraphics[height=4 cm,origin=c,angle=0]{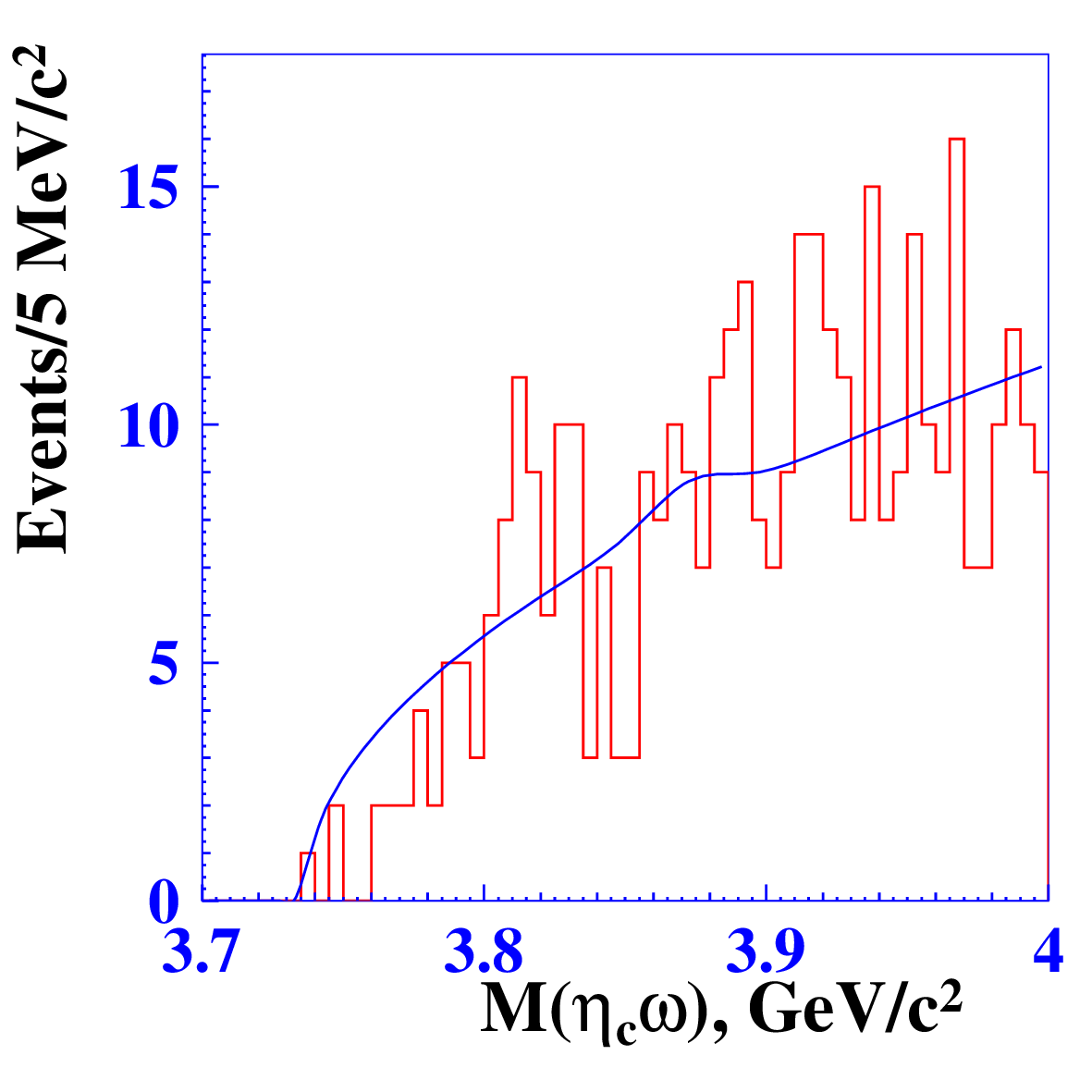}
\caption{The distributions of $\eta_c\pi^+\pi^-$ (left) and $\eta_c\omega$ (right) invariant masses in the search for the $X_1(3872)$. The threshold of the $\eta_c\omega$ invariant mass is slightly shifted relative to the sum of the $\omega$ and $\eta_c$ nominal masses due to the mass windows of these two mesons, as confirmed by MC studies.}
\label{pic:2}
\end{figure}

For the ($\pi^+\pi^-$) mode, we perform a maximum-likelihood fit of the $\eta_c\pi^+\pi^-$ invariant mass distribution that is, again, modeled by a peaking signal and a smooth background; the fit function is
\begin{equation}
f(x)=N_s\left[\alpha G(M,\sigma_1)+(1-\alpha)G(M,\sigma_2)\right]+c_0+c_1x.
\label{eq:5}
\end{equation}
For each intermediate resonance, we generate signal MC that incorporates the corresponding ``$X(3872)$-like'' particle quantum numbers. From the signal-MC fit, we fix the mean value $M$, the standard deviations $\sigma_1$ and $\sigma_2$, and the Gaussian fraction $\alpha$; in addition, we obtain the detection efficiency for the mode. From the fit, we obtain the signal yield $N_s$, which is shown in table~\ref{tab:11}. The $\eta_c\pi^+\pi^-$ invariant mass distribution is shown in figure~\ref{pic:2} (left). 

We validate our procedure by applying it to the decay $B^{\pm}\to K^{\pm}\psi(2S)$, $\psi(2S)\to J/\psi\pi^+\pi^-$. This decay is similar to the ($\pi^+\pi^-$) decay except that we reconstruct the $\psi(2S)$ meson in place of the $X_1(3872)$ and the $J/\psi$ in place of the $\eta_c$. The $J/\psi$ meson, like the $\eta_c$, is reconstructed via the $K^0_SK^{\pm}\pi^{\mp}$ final state. The selection criteria for this decay are the same as for the ($\pi^+\pi^-$) decay except for the invariant mass of the $K^0_SK^{\pm}\pi^{\mp}$ combination: $3.077$ GeV/$c^2$ $<M(K^0_SK^{\pm}\pi^{\mp})<3.117$ GeV/$c^2$. The mean number of $B$
candidates per event is 1.7. In case of multiple $B$ candidates, the one with the minimum differences for
$|m_{K^0_S}-M(\pi^+\pi^-)|$ and $|m_{J/\psi}-M(K^0_SK^{\pm}\pi^{\mp})|$ and the best
vertex coordinate is chosen.
We fit the $J/\psi\pi^+\pi^-$ invariant mass distribution and obtain the number of signal events $N_s=20.2\pm 6.5$, which corresponds to a significance of 3.5 standard deviations ($\sigma$). The significance is estimated using the value of $-2\sqrt{\ln{\frac{\mathcal{L}_{\rm 0}}{\mathcal{L}_{\rm max}}}}$, where ${\mathcal{L}_{\rm max}}$ (${\mathcal{L}_{\rm 0}}$) denotes the likelihood value when the yield is allowed to vary (is set to zero).
The expected number of events estimated using the world averages of the known branching fractions~\cite{PDG} is $22\pm 4$, which is consistent with $N_s$.

In the analysis of the ($\omega$) mode, we use the sum of two Gaussians to describe the signal and a threshold square-root function to describe the background:
\begin{equation}
f(x)=N_s\left[\alpha G(M,\sigma_1)+(1-\alpha)G(M,\sigma_2)\right]+c_0\sqrt{x-c_1}.
\label{eq:6}
\end{equation}
The $\eta_c\omega$ invariant mass distribution is shown in figure~\ref{pic:2} (right). 

\begin{figure}[htb]
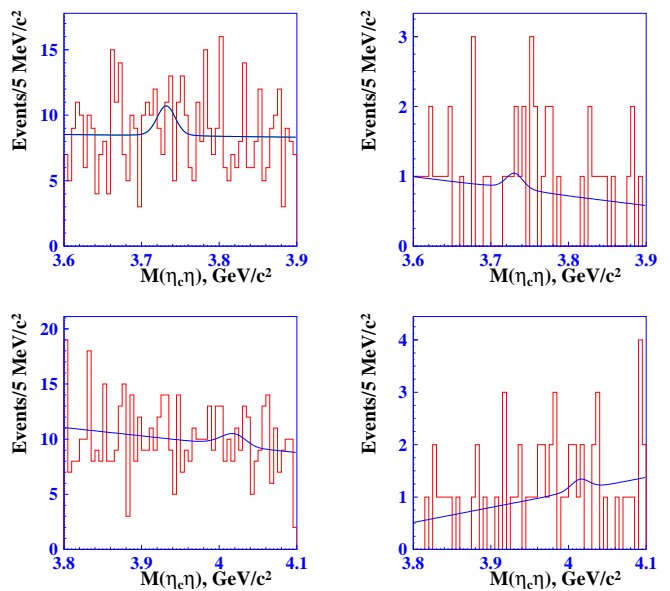

\centering
\includegraphics[height=4 cm]{fig6a.eps}
\hfill
\includegraphics[height=4 cm,origin=c,angle=0]{fig6b.eps}
\vfill
\includegraphics[height=4 cm]{fig6c.eps}
\hfill
\includegraphics[height=4 cm,origin=c,angle=0]{fig6d.eps}
\caption{The combined fit projections of the $\eta_c\eta$ invariant mass distributions in case of the $\eta\to\gamma\gamma$ (left) and $\eta\to\pi^+\pi^-\pi^0$ (right) modes corresponding to the search for the $X(3730)$ (top) and $X(4014)$ (bottom) resonances.}
\label{pic:3}
\end{figure}

In the $X(3730)$ mass region of the ($\eta$) mode, the fitting function is
\begin{eqnarray}
f_i(x)&=&N_{\rm eff}\varepsilon_{i}{\mathcal B}_{i}\left[\alpha_iG(M_{i},\sigma_{1,i})+(1-\alpha_i)G(M_{i},\sigma_{2,i})\right]+ \nonumber\\
& & c_{0,i}+c_{1,i}x,
\label{eq:9}
\end{eqnarray}
where  $N_{\rm eff}$ is the effective number of signal events and $i$ refers to either $\eta\to\gamma\gamma$ or $\eta\to\pi^+\pi^-\pi^0$ decay. 
The $\eta_c\eta$ invariant mass distributions in the $X(3730)$ mass region are shown in figure~\ref{pic:3} (top). 

In the $X(4014)$ mass region of the ($\eta$) mode, the fitting function is
\begin{equation}
f(x)=\left\{
\begin{array}{l}
N_{\rm eff}\varepsilon_{2\gamma}{\mathcal B}_{2\gamma}G_{\rm LG}(M_{2\gamma},\sigma_{2\gamma},P) + c_{0,1} + c_{1,1}x,\\
\\
N_{\rm eff}\varepsilon_{3\pi}{\mathcal B}_{3\pi}[\alpha G(M_{3\pi},\sigma_{1,3\pi})+ \\
(1-\alpha)G(M_{3\pi},\sigma_{2,3\pi})]+c_{0,2}+c_{1,2}x.
\end{array} \right.
\label{eq:10}
\end{equation}
The $\eta_c\eta$ invariant mass distributions in the $X(4014)$ mass region are shown in figure~\ref{pic:3} (bottom). 

\begin{figure}[htb]
\centering
\includegraphics[height=4 cm]{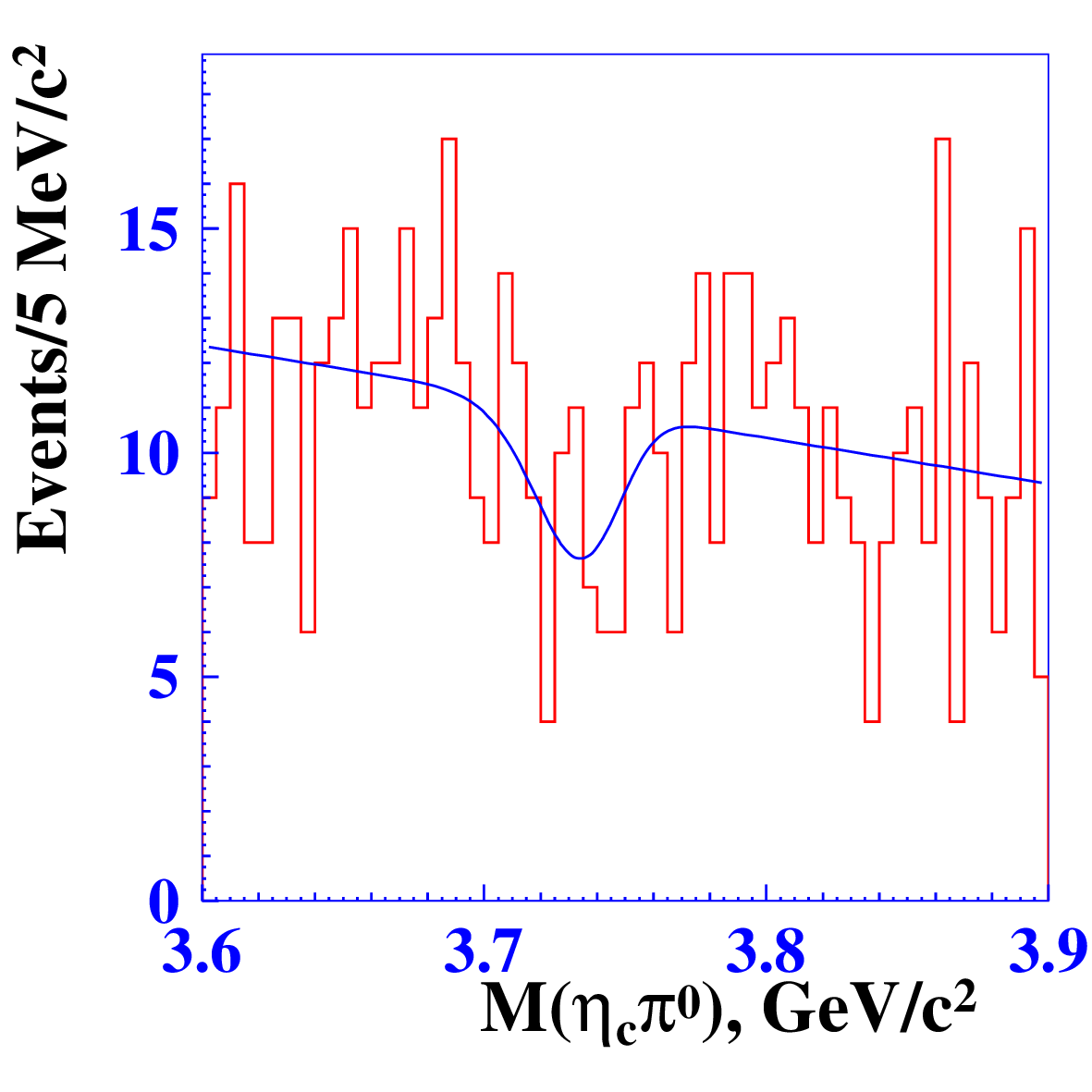}
\hfill
\includegraphics[height=4 cm,origin=c,angle=0]{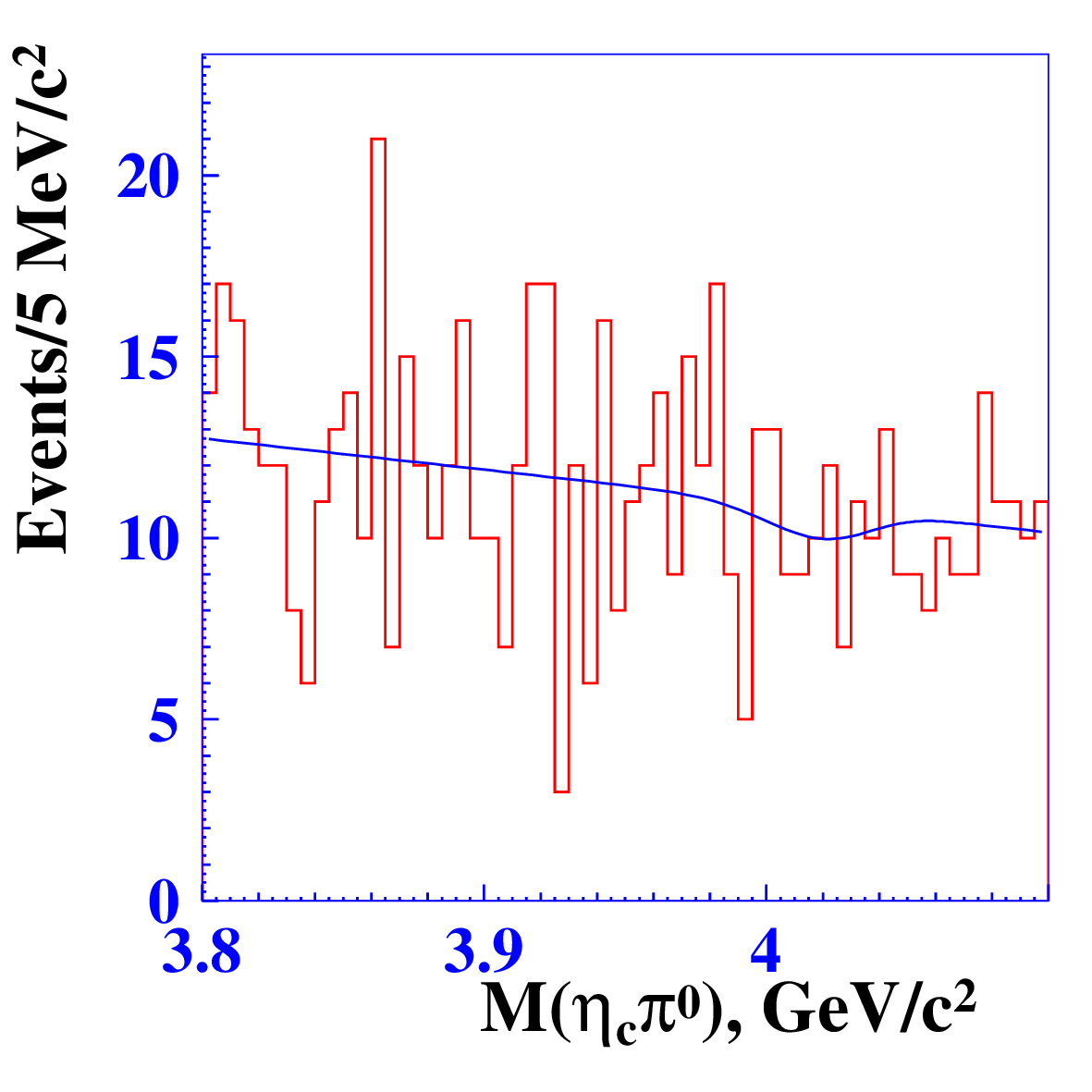}
\caption{The $\eta_c\pi^0$ invariant mass distributions corresponding to the search for the $X(3730)$ (left) and $X(4014)$ (right) resonances.}
\label{pic:4}
\end{figure}

For the ($\pi^0$) mode, we use
\begin{equation}
f(x)=N_sG_{\rm LG}(M,\sigma,P)+c_0+c_1x.
\label{eq:11}
\end{equation}
The $\eta_c\pi^0$ invariant mass distributions are shown in figure~\ref{pic:4}. 

The fit results are summarized in table~\ref{tab:11}.
\begin{table}[htb]
\centering
\begin{tabular}{|c||c|c|c|} 
\hline
Decay mode & Fit. function & Efficiency, \% & Yield \\
\hline \hline
$X_1(3872)\to \eta_c\pi^+\pi^-$ & (\ref{eq:5}) & $7.95\pm 0.02$ &
$17.9\pm 16.5$ \\
\hline
$X_1(3872)\to\eta_c\omega$ & (\ref{eq:6}) & $1.92\pm 0.02$ & $6.0\pm 12.5$\\
\hline
$X(3730)\to\eta_c\eta$, & & & \\
\cline{3-4}
$\eta\to\gamma\gamma$ & (\ref{eq:9}) & $6.57\pm 0.02$ & $13.8\pm 9.9$ \\
\cline{3-4}
$\eta\to\pi^+\pi^-\pi^0$ &  & $ 1.18\pm 0.01$ & $1.4\pm 1.0$ \\
\hline
$X(3730)\to\eta_c\pi^0$ & (\ref{eq:11}) & $6.52\pm 0.02$ & $-25.6\pm 10.4$ \\
\hline
$X(4014)\to\eta_c\eta$, & & & \\
\cline{3-4}
$\eta\to\gamma\gamma$ & (\ref{eq:10}) & $7.09\pm 0.02$ & $8.9\pm 11.0$ \\
\cline{3-4}
$\eta\to\pi^+\pi^-\pi^0$ &  & $1.78\pm 0.01$ & $1.3\pm 1.6$ \\
\hline
$X(4014)\to\eta_c\pi^0$ & (\ref{eq:11}) & $7.55\pm 0.02$ & $-8.1\pm 13.2$ \\
\hline
\end{tabular}
\caption{\label{tab:11} Fit results for the $X_1(3872)$, $X(3730)$ and $X(4014)$ resonances.}
\end{table}

\subsection{$Z(3900)^0$ and $Z(4020)^0$}

\begin{figure}[htb]
\centering
\includegraphics[height=4 cm]{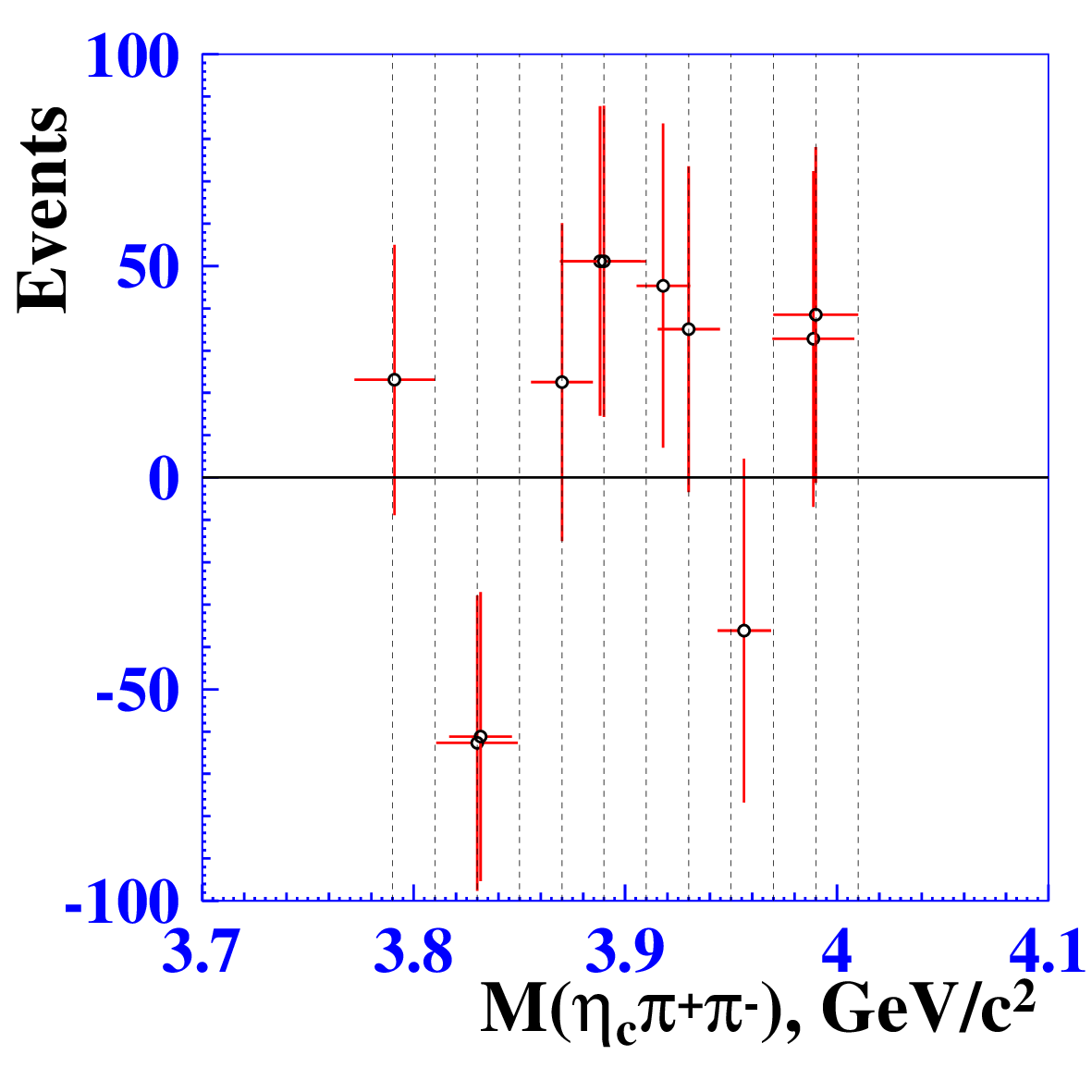}
\hfill
\includegraphics[height=4 cm,origin=c,angle=0]{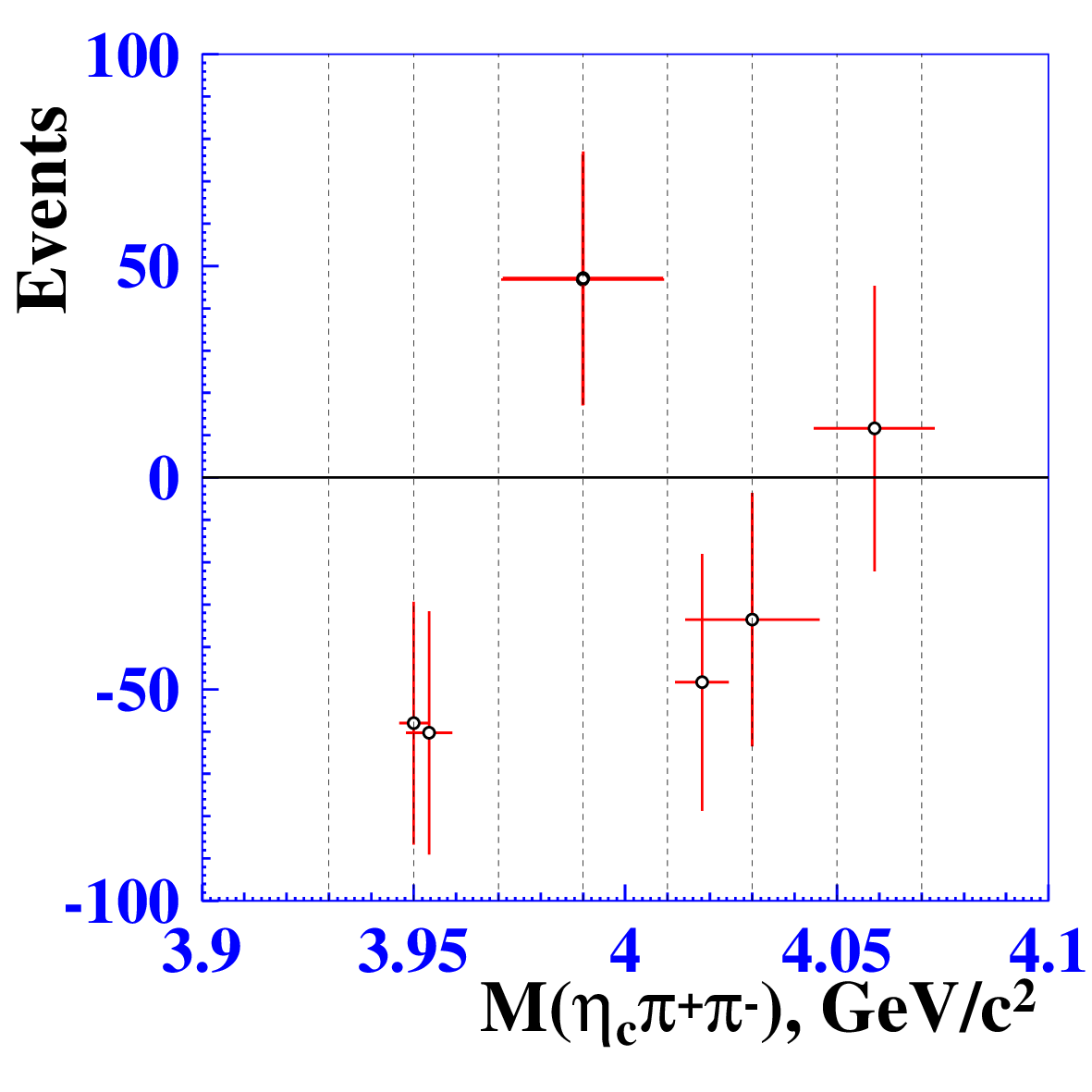}
\caption{Dependence of the signal yield of the $Z(3900)^0$ (left) and $Z(4020)^0$ (right) on the mass bin. The mass bin is a $20$ MeV/$c^2$ window to which the mass is confined and scanned in $20$ MeV/$c^2$ steps across the fit range.}
\label{pic:33}
\end{figure}

\begin{figure}[htb]
\centering
\includegraphics[height=4 cm]{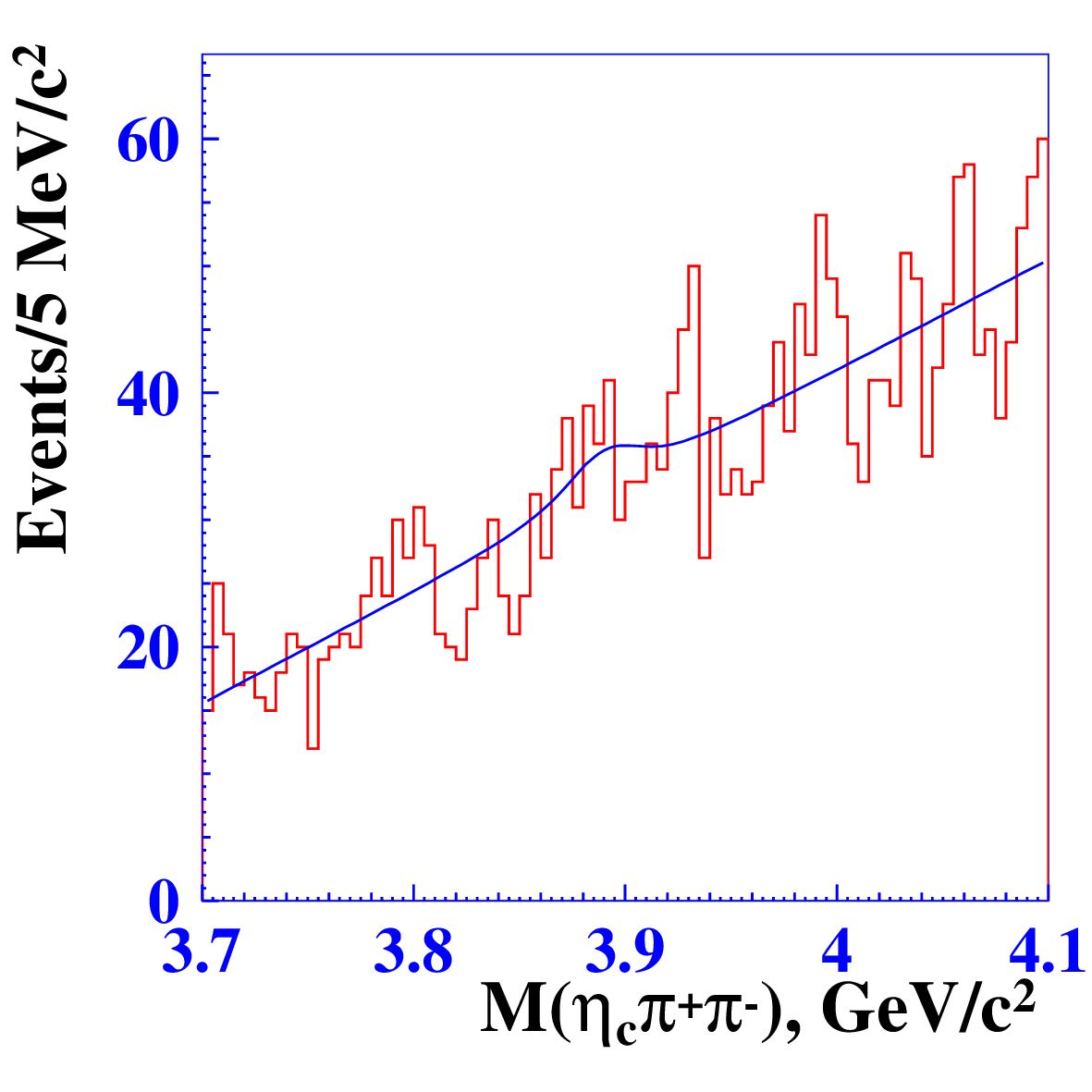}
\hfill
\includegraphics[height=4 cm,origin=c,angle=0]{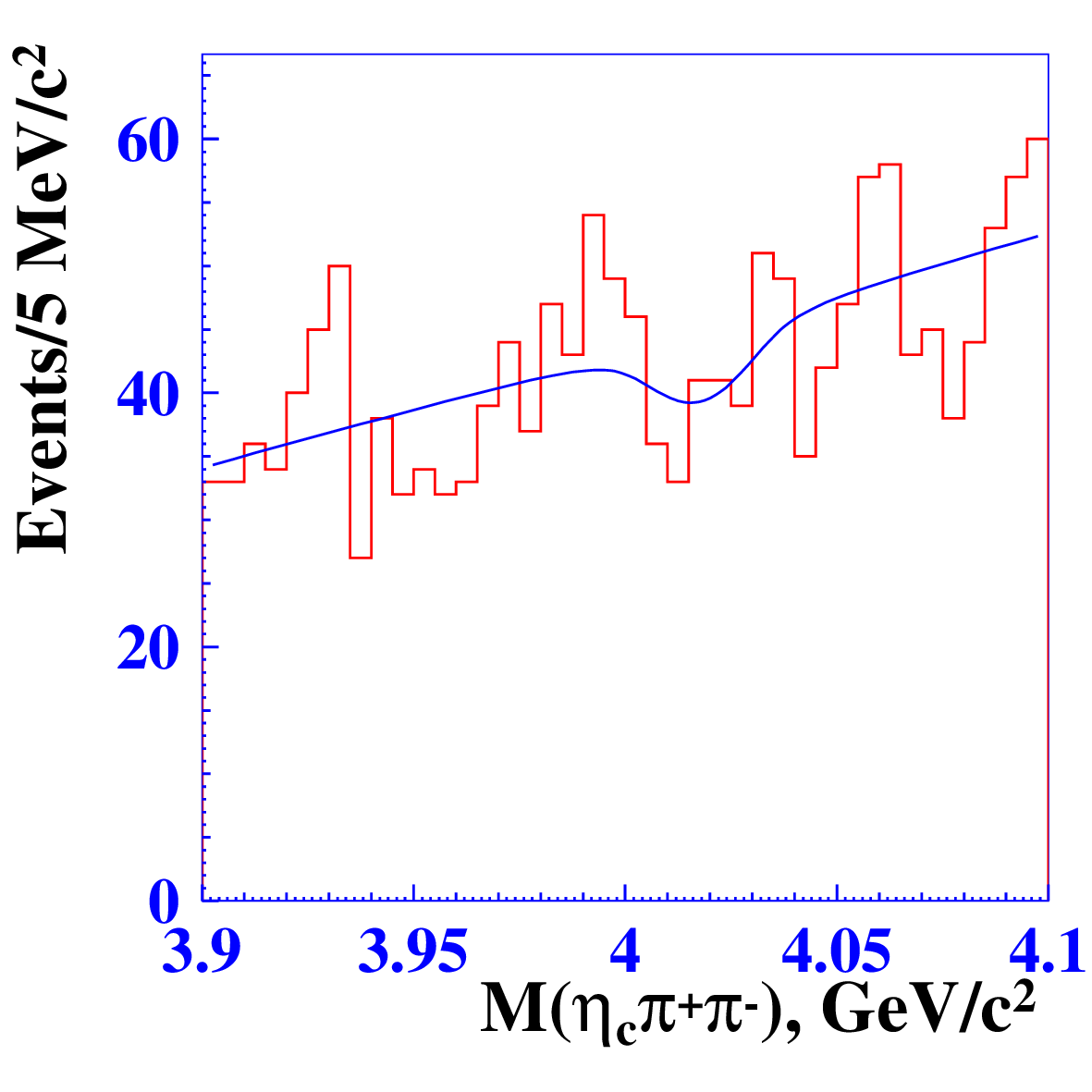}
\caption{Examples of the fit within the mass bin containing $3.9$ GeV/$c^2$ for $Z(3900)^0$ (left) and $4.02$ GeV/$c^2$ for $Z(4020)^0$ (right).}
\label{pic:33a}
\end{figure}

We perform a sequence of binned maximum likelihood fits of the $\eta_c\pi^+\pi^-$ invariant mass using the convolution of a Breit-Wigner and a Gaussian for the signal and a linear polynomial for the background:
\begin{equation}
f(x)=b(M,\Gamma)\otimes G(0,\sigma_{\rm res})+c_0+c_1x.
\label{eq:z}
\end{equation}
The Gaussian models the detector resolution, which is assumed to be similar to that obtained in Ref.~\cite{Vinok} and equal to $9.8$ MeV/$c^2$. The Breit-Wigner mass is confined to a $20$ MeV/$c^2$ window (the so-called mass bin) that is scanned in $20$ MeV/$c^2$ steps across the range $(3.79-4.01)$ GeV/$c^2$ for the $Z(3900)^0$ and $(3.93-4.07)$ GeV/$c^2$ for the $Z(4020)^0$. The width is fixed to the weighted mean of the previously measured values ($35$ MeV/$c^2$ for the $Z(3900)^0$ and $12$ MeV/$c^2$ for the $Z(4020)^0$). The background is described by a linear polynomial. The detection efficiency is obtained from signal MC: $(9.64 \pm 0.03)$\% for the $Z(3900)^0$ and $(10.42 \pm 0.03)$\% for the $Z(4020)^0$. The obtained signal yield is shown in figure~\ref{pic:33}. The examples of the fit within the mass bin containing $3.9$ GeV/$c^2$ for $Z(3900)^0$ and $4.02$ GeV/$c^2$ for $Z(4020)^0$ are shown in figure~\ref{pic:33a}. 

\subsection{$X(3915)$}

\begin{figure}[htb]
\centering
\includegraphics[height=4 cm]{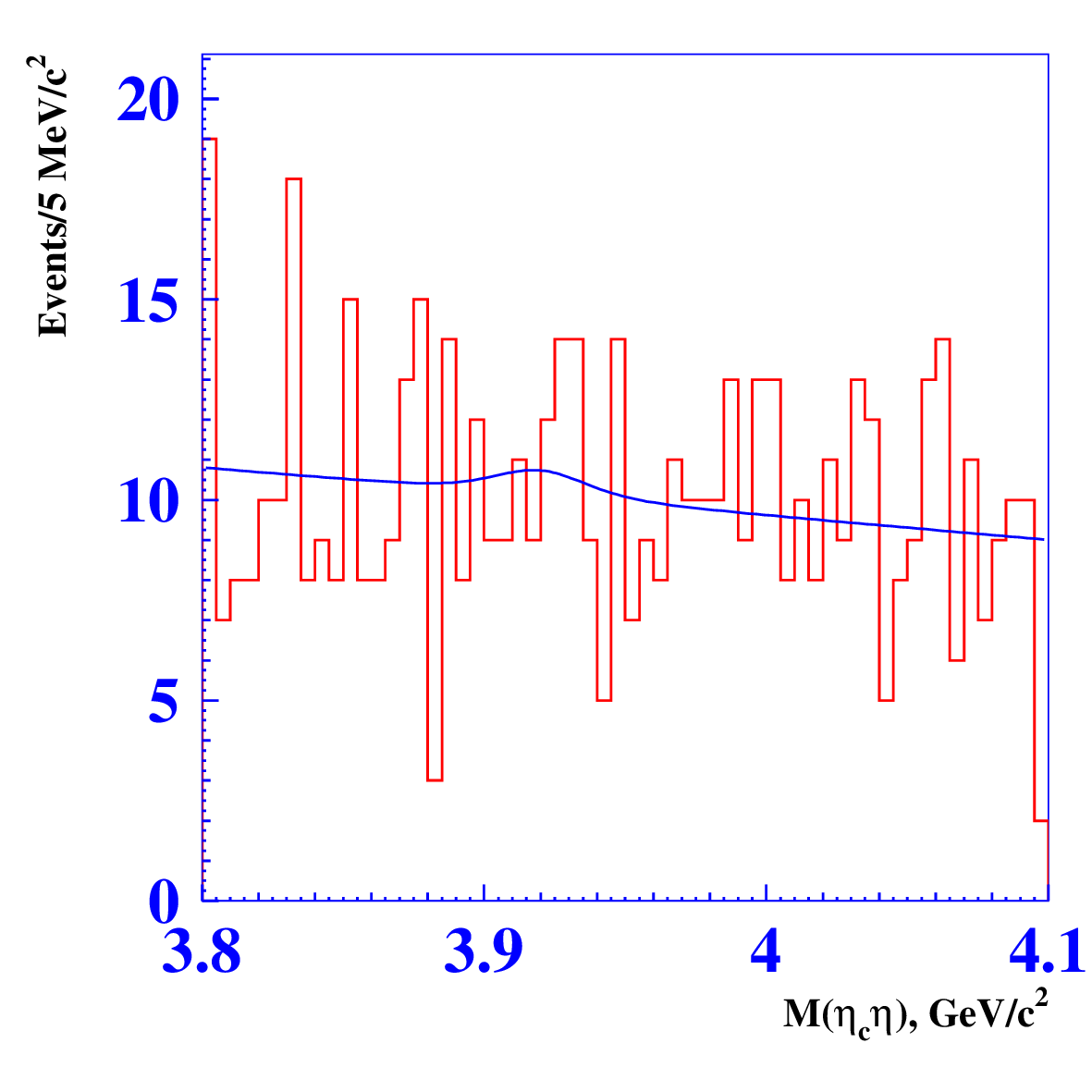}
\hfill
\includegraphics[height=4 cm,origin=c,angle=0]{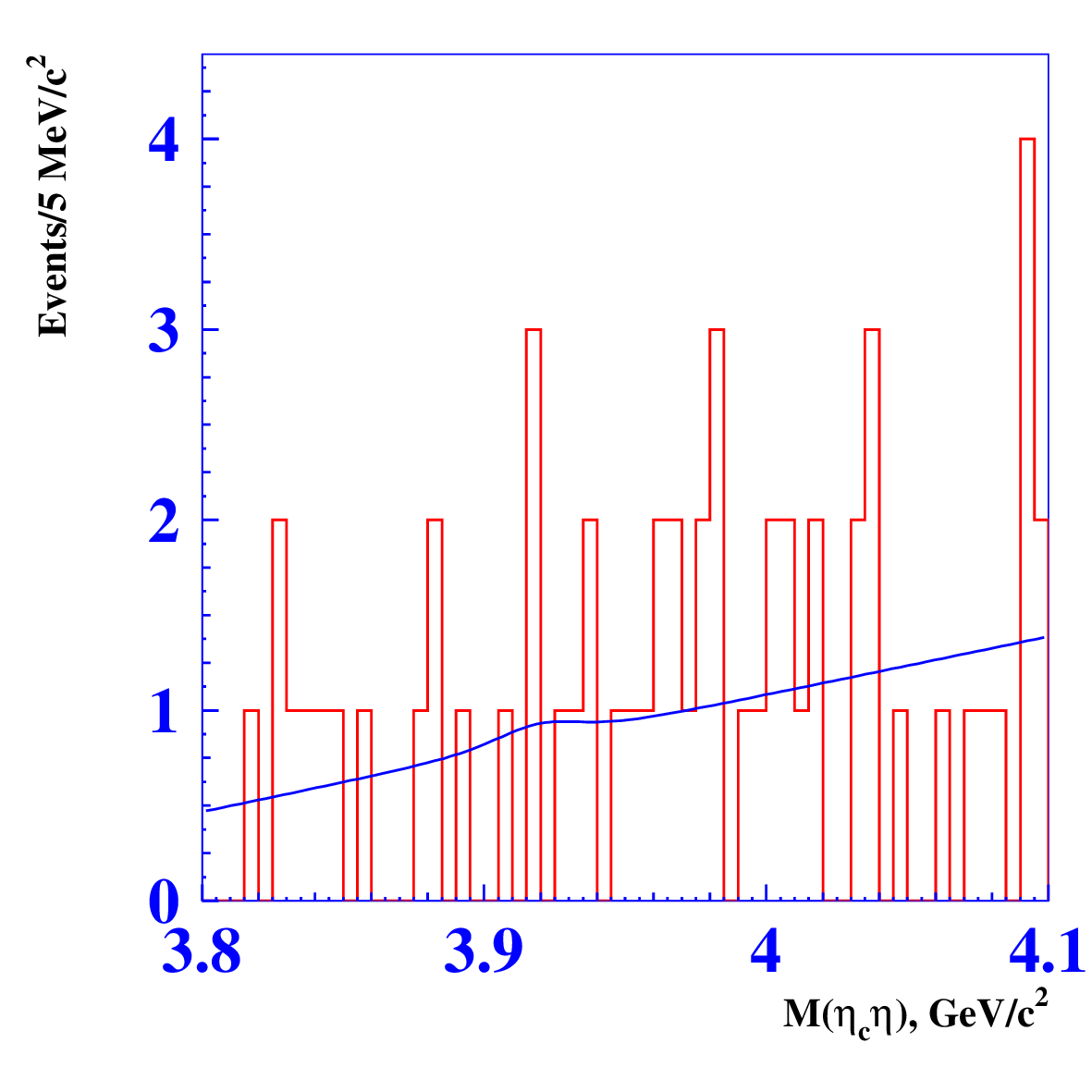}
\caption{The combined fit projections of the $\eta_c\eta$ invariant mass distributions in case of the $\eta\to\gamma\gamma$ (left) and $\eta\to\pi^+\pi^-\pi^0$ (right) modes corresponding to the search for the $X(3915)$ resonance.}
\label{pic:39151}
\end{figure}

For the $\eta_c\eta$ invariant mass distribution, we perform a combined fit of two decay modes of the $\eta$ meson:
\begin{eqnarray}
f_i(x)&=&N_{\rm eff}\varepsilon_{i}{\mathcal B}_{i}\left[b(M_i,\Gamma_i)\otimes G(0,\sigma_i)\right] + \nonumber\\
& & c_{0,i}+c_{1,i}x,
\label{eq:x39151}
\end{eqnarray}
where  $N_{\rm eff}$ is the effective number of signal events and $i$ refers to either $\eta\to\gamma\gamma$ or $\eta\to\pi^+\pi^-\pi^0$ decay. The detector resolution $\sigma_i$ is obtained from signal MC, taking into account the resolution degradation and is equal to $13.6$ MeV/$c^2$ for $\eta\to\gamma\gamma$ and $12.3$ MeV/$c^2$ for $\eta\to\pi^+\pi^-\pi^0$. The corresponding $\eta_c\eta$ invariant mass distribution is shown in figure~\ref{pic:39151}.

\begin{figure}[htb]
\centering
\includegraphics[height=4 cm]{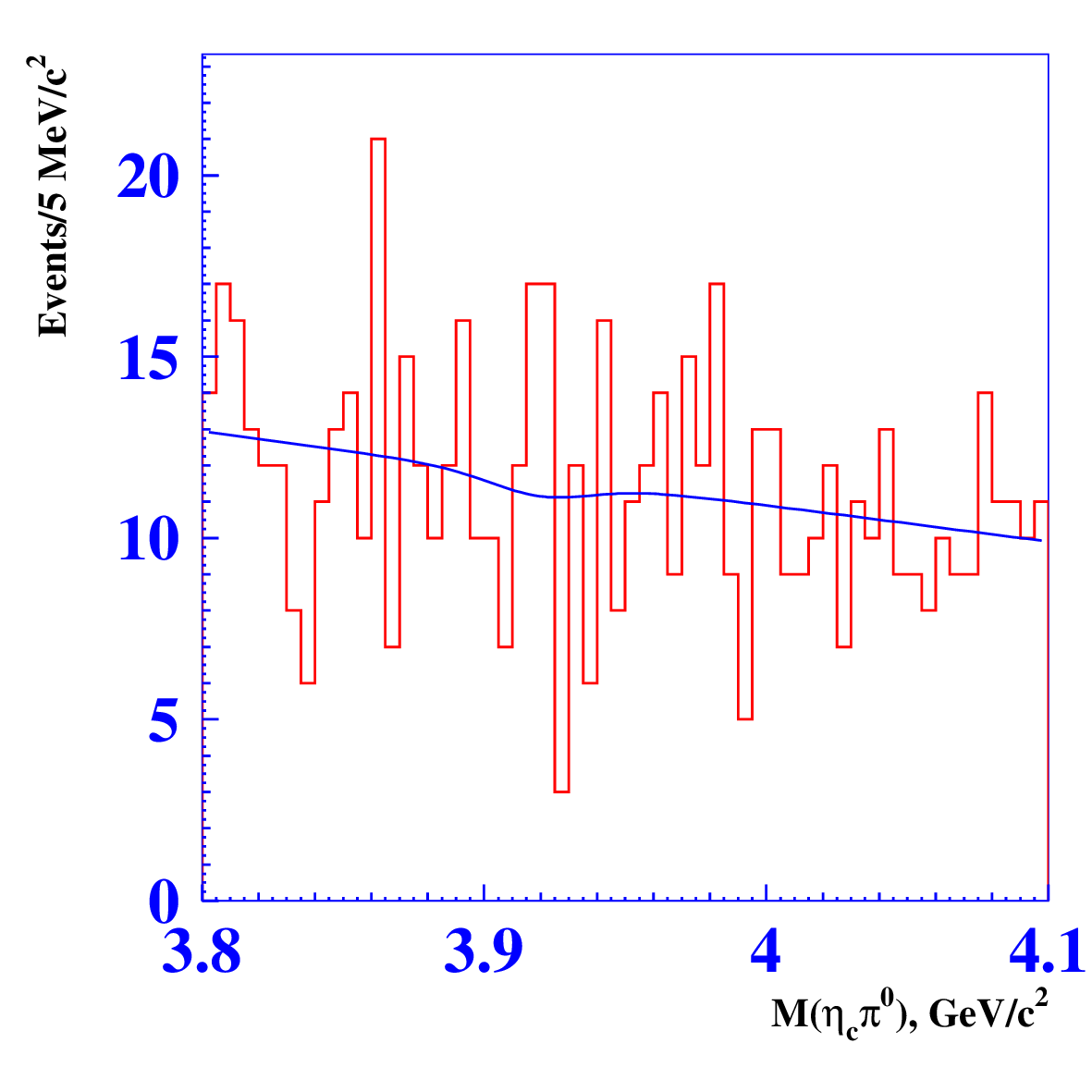}
\caption{The $\eta_c\pi^0$ invariant mass distribution corresponding to the search for the $X(3915)$ resonance.}
\label{pic:39152}
\end{figure}

We fit the $\eta_c\pi^0$ invariant mass distribution with the function in Eq.~(\ref{eq:z}). The detector resolution is obtained from signal MC, taking into account the resolution degradation, and is equal to $15.7$ MeV/$c^2$. The corresponding $\eta_c\pi^0$ invariant mass distribution is shown in figure~\ref{pic:39152}.

The fit results are summarized in table~\ref{tab:51}.
\begin{table}[htb]
\centering
\begin{tabular}{|c||c|c|c|} 
\hline
Decay mode & Fit. function & Efficiency, \% & Yield \\
\hline \hline
$X(3915)\to\eta_c\eta$, & & & \\
\cline{3-4}
$\eta\to\gamma\gamma$ & (\ref{eq:x39151}) & $6.60\pm 0.02$ & $7.1\pm 14.5$ \\
\cline{3-4}
$\eta\to\pi^+\pi^-\pi^0$ &  & $ 1.64\pm 0.01$ & $1.0\pm 2.1$ \\
\hline
$X(3915)\to\eta_c\pi^0$ & (\ref{eq:z}) & $6.88\pm 0.02$ & $-6.9\pm 17.3$ \\
\hline
\end{tabular}
\caption{\label{tab:51} Fit results for the $X(3915)$ resonance.}
\end{table}

\section{Systematic uncertainties}

Systematic uncertainties are categorized as follows:
\begin{enumerate}
\item Additive systematic uncertainties affect the number of signal events and are estimated by the variation of the fit conditions. They are displayed as the numbers of events in tables~\ref{tab:4} and~\ref{tab:2} and arise from the sources listed below.
{\it (i)} To obtain the error related to the resolution degradation, we vary the corrected variance of the fitting function within its statistical uncertainty. 
{\it (ii)} We assume that the combinatorial background can be parameterized with a 
first-order polynomial. To obtain the background shape uncertainty, 
we describe the background by a second-order polynomial and compare the results. For the ($\omega$) mode, we change the square root to the fourth root. 
{\it (iii)} To estimate the systematic error associated with the selection criteria, we relax the criteria on $\Delta E$, $M_{\rm bc}$, and the invariant masses of $\eta_c$, $\omega$ and $\eta$ by 50\%.
{\it (iv)} We vary the bin size between 2.5 to 7.5 MeV/c$^2$ and determine the corresponding systematic error.
\item Multiplicative systematic uncertainties affect the product branching fractions. They are displayed in percent in table~\ref{tab:3} and arise from the sources listed below.
{\it (i)} The number of $B\bar{B}$ pairs is calculated from the difference between the 
number of hadronic events on resonance and the scaled number of those 
off resonance. The systematic error is dominated by the uncertainty in the scale factor and is equal to $\sim$1.4\%~\cite{BB}.
{\it (ii)} The uncertainties on the $\omega$, $\pi^0$, $\eta$, $\eta_c$ and $K^0_S$ decay branching fractions are taken from Ref.~\cite{PDG}.
{\it (iii)} The statistical error of the efficiency determined by the signal MC is also taken as a systematic uncertainty. For $Z(3900)^0$ and $Z(4020)^0$ decays, we take into account the efficiency variation in the $\eta_c\rho$ and $\eta_cf_0$ decay modes. For the decays $B^{\pm}\to K^{\pm}\eta_c\pi^+\pi^-$ and $B^{\pm}\to K^{\pm}\eta_c\pi^0$, we assume alternative decay modes containing intermediate $K^{*0}$, $K^*(1410)^{\pm}$ and $\rho$ mesons, and take into account the difference of the MC detection efficiency as the corresponding systematic uncertainty.
{\it (iv)} An analysis of the charged track reconstruction
uncertainty as a function of particle momentum gives an estimate of 0.34\% per charged track.
{\it (v)} To determine the errors due to $K$ and $\pi$ meson identification, data from the
 analysis of the process $D^{*+}\to D^0\pi^+$ followed by the decay
$D^0\to K^-\pi^+$ are used. The
uncertainty in $K^{\pm}$ identification is 0.8\% per $K$ meson and 
the corresponding value for $\pi^{\pm}$ identification is 0.5\% per $\pi$ meson.
{\it (vi)} Estimation of the contribution of the $\eta$ and $\pi^0$ reconstruction uncertainty is
carried out using the comparison of the number of reconstructed
$\eta\to3\pi^0$ and $\eta\to\gamma\gamma$ events. Such an estimate gives
2\% per $\eta$ and $\pi^0$ meson.
{\it (vii)} The contribution of the $K^0_S$ reconstruction uncertainty is estimated to be 4.4\%~\cite{Ks}.
{\it (viii)} We also take into account the deviation of MC from the data by applying a correction
to the efficiency:
$\varepsilon_{\rm Data}/\varepsilon_{\rm MC}$ is 0.9996 for each kaon
and 0.9756 for each pion.
\end{enumerate}

For the ($\eta$) mode, the $\eta$ meson is reconstructed 
in two decay modes: $\eta\to\gamma\gamma$ and $\eta\to\pi^+\pi^-\pi^0$. 
To estimate the systematic uncertainty, we compare the systematic errors given by each $\eta$ decay mode and take the maximum as the uncertainty for the reconstruction.  

For the $Z(3900)^0$ and $Z(4020)^0$ decays to $\eta_c\pi^+\pi^-$, the additive systematic uncertainty is assumed to be the same as in $X_1(3872)\to\eta_c\pi^+\pi^-$. In addition, we vary the resonance width in the $(15-65)$ MeV/$c^2$ interval for the $Z(3900)^0$ and $(2.5-27.5)$ MeV/$c^2$ interval for the $Z(4020)^0$. The intervals are chosen according to the variation of the previously measured $Z(3900)^{\pm}$ and $Z(4020)^{\pm}$ width values. The total  additive systematic uncertainty is $28.5$ events for the $Z(3900)^0$ and $25.9$ events for the $Z(4020)^0$.

For the $X(3915)$ decays, the systematic uncertainty is taken from the similar $X(4014)$ decays (see tables~\ref{tab:2} and~\ref{tab:3}).

\begin{table}[htb]
\centering
\begin{tabular}{|c||c|c|c|c|} 
\hline
Source & $\eta_c\pi^+\pi^-$ & $\eta_c\omega$ & $\eta_c\eta$ & $\eta_c\pi^0$ \\
\hline \hline
\scriptsize{Background parameterization} & $1$ & $44$ & $2687$ & $2$\\
\scriptsize{Selection criteria} & $<1$ & $<1$ & $1695$ & $33$\\
\scriptsize{Bin size} & $18$ & $2$ & $430$ & $9$\\
\hline \hline
Total (events) & $18$ & $44$ & $3206$ & $34$\\
\hline
\end{tabular} 
\caption{\label{tab:4} Additive systematic uncertainties (in events) for $B$ decays without intermediate resonances. For the $\eta_c\eta$ mode, the uncertainty corresponds to the effective number of events $N_{\rm eff}$.}
\end{table}
\begin{table}[htb]
\centering
\begin{tabular}{|c||c|c|c|c|c|c|} 
\hline
Source & $\eta_c\pi^+\pi^-$ & $\eta_c\omega$ & \multicolumn{2}{c|}{$\eta_c\eta$} & \multicolumn{2}{c|}{$\eta_c\pi^0$} \\
\hline
Mass, MeV/c$^2$ & $3872$ & $3872$ & $3730$ & $4014$ & $3730$ & $4014$ \\
\hline \hline
\scriptsize{Resolution degradation} & $1.2$ & $<1$ & $68$ & $28$ & $1.4$ & $0.2$\\ 
\scriptsize{Background parameterization} & $5.8$ & $3.6$ & $18$ & $8$ & $0.8$ & $0.3$\\
\scriptsize{Selection criteria} & $23.9$ & $5.4$ & $293$ & $280$ & $5.2$ & $9.3$\\
\scriptsize{Bin size} & $1.2$ & $7.7$ & $30$ & $71$ & $2.4$ & $4.4$\\
\hline \hline
Total (events) & $24.7$ & $10.1$ & $303$ & $290$ & $5.9$ & $10.3$\\
\hline
\end{tabular}
\caption{\label{tab:2} Additive systematic uncertainties (in events) for $B$ decays containing the $X(3872)$-like particles. For the $\eta_c\eta$ mode, the uncertainty corresponds to the effective number of events $N_{\rm eff}$.}
\end{table}
\begin{table}[htb]
\centering
\begin{tabular}{|c||c|c|c|c|} 
\hline
Source & $\eta_c\pi^+\pi^-$ & $\eta_c\omega$ & $\eta_c\eta$ & $\eta_c\pi^0$ \\
\hline \hline
Number of $B\bar{B}$ pairs & $1.4$ & $1.4$ & $1.4$ & $1.4$\\ 
${\mathcal B}(\omega\to\pi^+\pi^-\pi^0)$ & --- & $0.8$ & --- & ---\\
${\mathcal B}(\pi^0\to\gamma\gamma)$ & --- & $<0.1$ & $<0.1$ & $<0.1$\\
${\mathcal B}(\eta\to\gamma\gamma)$ & --- & --- & $0.5$ & ---\\
${\mathcal B}(\eta\to\pi^+\pi^-\pi^0)$ & --- & --- & $1.2$ & ---\\
${\mathcal B}(\eta_c\to K^0_SK^{\pm}\pi^{\mp})$ & $6.8$ & $6.8$ & $6.8$ & $6.8$\\
${\mathcal B}(K^0_S\to\pi^+\pi^-)$ & $0.1$ & $0.1$ & $0.1$ & $0.1$ \\
\hline
MC detection efficiency  &  &  &  & \\
\scriptsize{no resonance} & $35.8$ & $2.4$ & $1.3$ & $19.5$\\
\scriptsize{$X(3872)$-like} & $0.3$ & $0.5$ & $0.7$ & $0.3$ \\
\scriptsize{$Z(3900)^0$/$Z(4020)^0$} & $13.3$/$4.4$ & --- & --- & --- \\
\hline
Track reconstruction & $1.7$ & $1.7$ & $1.7$ & $1.0$  \\
$K^{\pm}$ identification & $1.6$ & $1.6$ & $1.6$ & $1.6$ \\
$\pi^{\pm}$ identification & $1.5$ & $1.5$ & $1.5$ & $0.5$ \\
$\eta$ reconstruction & --- & --- & $2.0$ & --- \\
$\pi^0$ reconstruction & --- & $2.0$ & $2.0$ & $2.0$ \\ 
$K^0_S$ reconstruction & $4.4$ & $4.4$ & $4.4$ & $4.4$ \\ 
\hline \hline
Total (\%) & & & & \\
\scriptsize{no resonance} & $36.8$ & $9.3$ & $9.3$ & $21.3$\\
\scriptsize{$X(3872)$-like} & $8.7$ & $9.0$ & $9.2$ & $8.7$ \\
\scriptsize{$Z(3900)^0$/$Z(4020)^0$} & $15.9$/$9.7$ & --- & --- & ---\\
\hline
\end{tabular}
\caption{\label{tab:3} Multiplicative systematic uncertainties (in \%).}
\end{table}

\section{Results}

Our data sample does not permit us to measure the branching products of production and decay of the states listed above nor the $B$ decay branching fractions so we set upper limits instead, taking into account the statistical and systematic uncertainties.
The expression for the required branching value is $z=N_s/(N_B\varepsilon\mathcal{B}_i)$, where $N_s$ is the yield, $N_B$ is the number of $B\bar{B}$ pairs, $\varepsilon$ is the detection efficiency, and $\mathcal{B}_i$ is the product of the branching fractions of the intermediate resonances. This expression can be written as $z=xy$, where $x=N_s$ and $y=\frac{1}{N_{B\bar{B}}\varepsilon\mathcal{B}_i}$. The distribution of the numerator $N_s$ is assumed to be Gaussian with mean $\mu$ and standard deviation $\sigma_x=\sqrt{\sigma_{\rm stat}^2+\sigma_{\rm add.syst}^2}$. The distribution of the inverse of the denominator is also assumed to be Gaussian with mean $\nu=1/(N_B\varepsilon\mathcal{B}_i)$ and standard deviation $\sigma_y=\sigma_{\rm mult.syst}\nu$. Thus, the distribution of $z$ can be written in the following way:
\begin{equation}
F(z)=\int_{-\infty}^{+\infty}G(x,\mu,\sigma_x)G(\frac{z}{x},\nu,\sigma_y\nu)\frac{1}{|x|}dx.
\label{eq:70}
\end{equation}  
The 90\% confidence level (C.L.) upper limit $U$ on $z$ is defined by
\begin{equation}
\frac{\int_0^UF(z)dz}{\int_0^{\infty}F(z)dz}=0.9.
\label{eq:71}
\end{equation}

Upper limits on the branching fractions and products for all the studied decay modes are shown in tables~\ref{tab:6} and~\ref{tab:5}.
\begin{table}[htb]
\centering
\begin{tabular}{|c|c|} 
\hline
Decay mode & Upper limit (90\% C.L.)\\
\hline \hline
$B^{\pm}\to K^{\pm}\eta_c\pi^+\pi^-$ & $3.9\times 10^{-4}$\\
$B^{\pm}\to K^{\pm}\eta_c\omega$ & $5.3\times 10^{-4}$\\
$B^{\pm}\to K^{\pm}\eta_c\eta$, & $2.2\times 10^{-4}$\\
$B^{\pm}\to K^{\pm}\eta_c\pi^0$ & $6.2\times 10^{-5}$\\ 
\hline
\end{tabular}
\caption{\label{tab:6} Results of branching fraction measurements for the $B$ decays without an intermediate resonance.}
\end{table}
\begin{table}[htb]
\centering
\begin{tabular}{|c|c|c|} 
\hline
Resonance & Decay mode & Upper limit (90\% C.L.)\\
\hline \hline
$X_1(3872)$ & $\eta_c\pi^+\pi^-$ & $3.0\times 10^{-5}$\\
 & $\eta_c\omega$ & $6.9\times 10^{-5}$\\
\hline
$X(3730)$ & $\eta_c\eta$ & $4.6\times 10^{-5}$\\
 & $\eta_c\pi^0$ & $5.7\times 10^{-6}$\\ 
\hline
 $X(4014)$ & $\eta_c\eta$ & $3.9\times 10^{-5}$\\
 & $\eta_c\pi^0$ & $1.2\times 10^{-5}$\\ 
\hline
\hline
$Z(3900)^0$ & $\eta_c\pi^+\pi^-$ & $4.7\times 10^{-5}$\\
\cline{1-1}
\cline{3-3}
$Z(4020)^0$ &  & $1.6\times 10^{-5}$\\
\hline
\hline
 $X(3915)$ & $\eta_c\eta$ & $4.7\times 10^{-5}$\\
 & $\eta_c\pi^0$ & $1.7\times 10^{-5}$\\ 
\hline
\end{tabular}
\caption{\label{tab:5} Results of branching fraction measurements for the $B$ decays containing an intermediate exotic resonance. For $Z(3900)^0$ and $Z(4020)^0$ resonances the results are shown under the assumption that the masses are close to those of their charged partners.}
\end{table}

\begin{figure}[htb]
\centering
\includegraphics[height=4 cm]{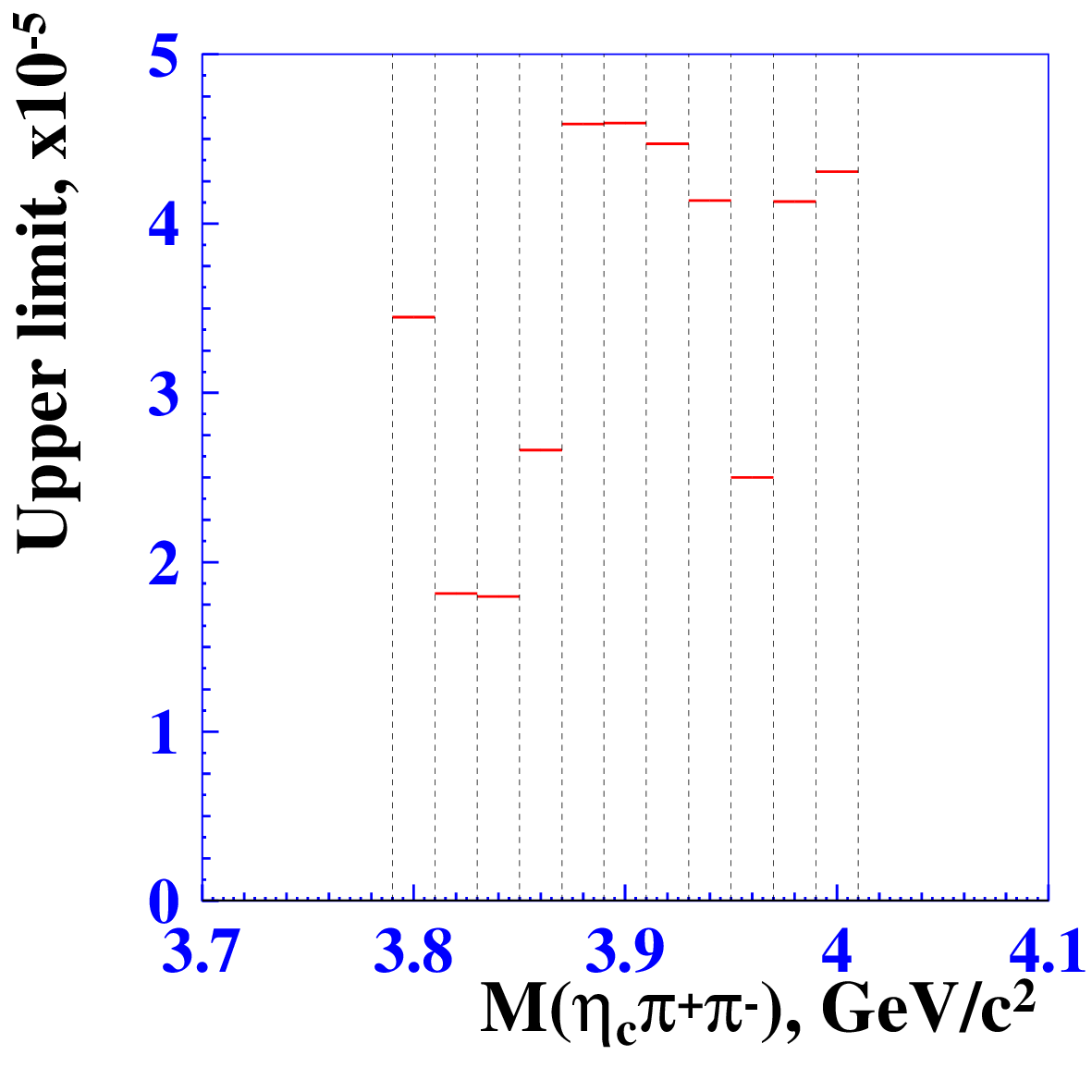}
\hfill
\includegraphics[height=4 cm,origin=c,angle=0]{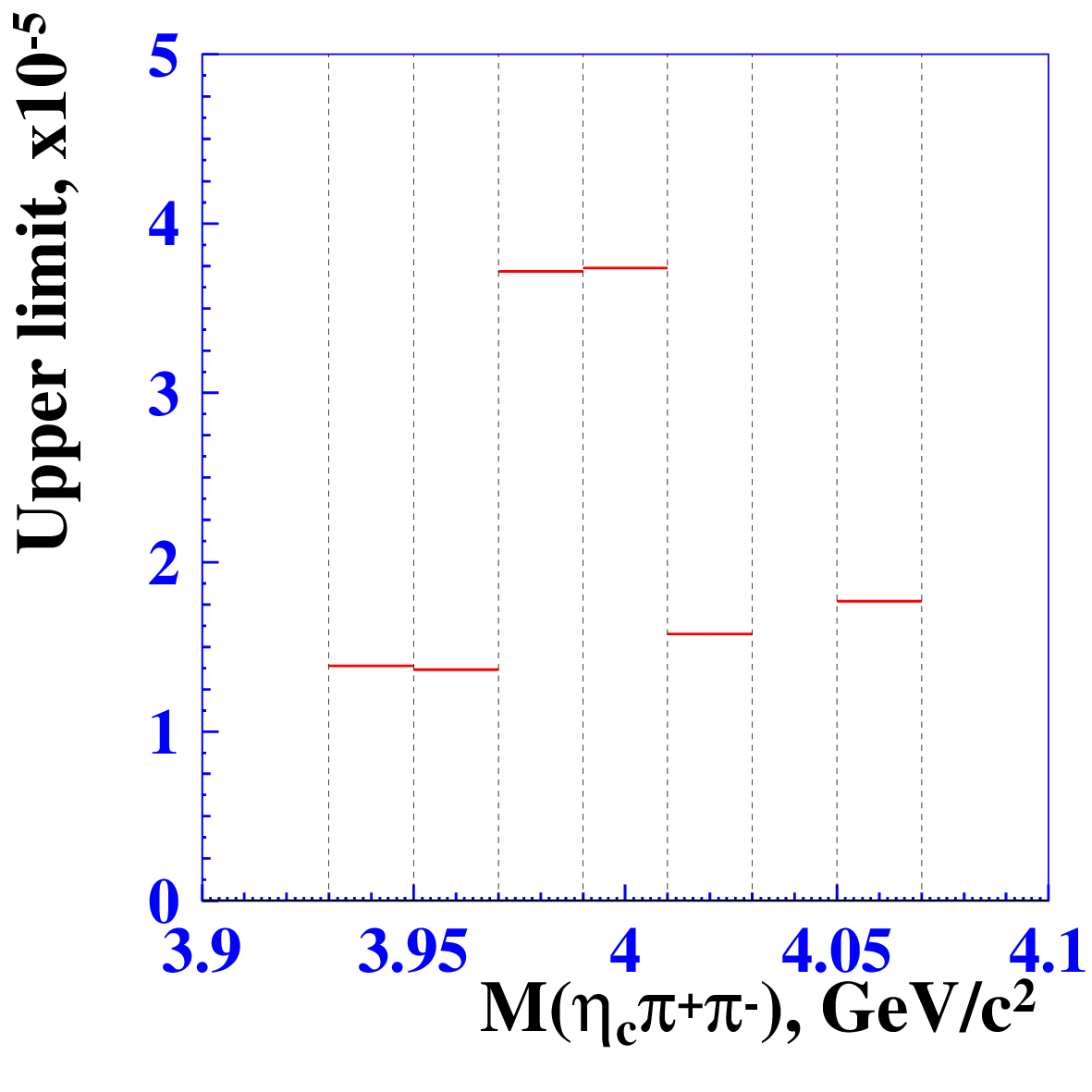}
\caption{Dependence on the mass bin of the upper limit for the $Z(3900)^0$ (left) and $Z(4020)^0$ (right) branching product.}
\label{pic:101}
\end{figure}

Figure~\ref{pic:101} shows the dependence on the mass bin of the upper limit for the branching product of the $Z(3900)^0$ and $Z(4020)^0$ production and decay: no significant signal is seen in any of the invariant mass bins. For the $Z(3900)^0$ resonance, we set upper limits in the range $(1.8-4.7)\times 10^{-5}$ for the mass region $(3.79-4.01)$ GeV/$c^2$. For the $Z(4020)^0$ resonance, we set upper limits in the range $(1.6-3.7)\times 10^{-5}$ for the mass region $(3.93-4.07)$ GeV/$c^2$. If we assume that the $Z(3900)^0$ and $Z(4020)^0$ masses are close to those of their charged partners, we obtain the upper limits on the product branching fractions shown in table~\ref{tab:5}.

Similarly, for the study of $X(3872)$-like particles, we performed a mass scan inside the fitting region, {\it i.e.,} a sequence of fits similar to the ones described above but with a mass floating in $20$ MeV/c$^2$-wide mass bins. No significant signal is found in any of the studied bins.

A similar analysis was performed by BaBar~\cite{BaBar3}, where the upper limit on the product $\sigma(\gamma\gamma\to X(3872))\times{\mathcal B}(X(3872)\to\eta_c\pi^+\pi^-)$ was set.  

In conclusion, we report a study of the following $B$ decays to final states with $\eta_c$: $B^{\pm}\to K^{\pm}\eta_c\pi^+\pi^-$, $B^{\pm}\to K^{\pm}\eta_c\omega$, $B^{\pm}\to K^{\pm}\eta_c\eta$ and $B^{\pm}\to K^{\pm}\eta_c\pi^0$.

We first study these $B$ decays without intermediate resonances and set 90\% C.L. upper limits on their branching fractions: ${\mathcal B}(B^{\pm}\to K^{\pm}\eta_c\pi^+\pi^-)<3.9\times 10^{-4}$, ${\mathcal B}(B^{\pm}\to K^{\pm}\eta_c\omega)<5.3\times 10^{-4}$, ${\mathcal B}(B^{\pm}\to K^{\pm}\eta_c\eta)<2.2\times 10^{-4}$ and ${\mathcal B}(B^{\pm}\to K^{\pm}\eta_c\pi^0)<6.2\times 10^{-5}$.

We then assume that the decays proceed through intermediate molecular states similar to the exotic $X(3872)$ particle, such as $X_1(3872)=D^0\bar{D}^{*0}-\bar{D}^0D^{*0}$, $X(3730)=D^0\bar{D}^0+\bar{D}^0D^0$ and $X(4014)=D^{*0}\bar{D}^{*0}+\bar{D}^{*0}D^{*0}$ and search for these states, setting 90\% C.L. upper limits on the following product branching fractions: ${\mathcal B}(B^{\pm}\to K^{\pm}X)\times{\mathcal B}(X\to\eta_c\pi^+\pi^-)<3.0\times 10^{-5}$ and ${\mathcal B}(B^{\pm}\to K^{\pm}X)\times{\mathcal B}(X\to\eta_c\omega)<6.9\times 10^{-5}$ for $X=X_1(3872)$, ${\mathcal B}(B^{\pm}\to K^{\pm}X)\times{\mathcal B}(X\to\eta_c\eta)<4.6\times 10^{-5}$ for $X=X(3730)$ and $<3.9\times 10^{-5}$ for $X=X(4014)$, ${\mathcal B}(B^{\pm}\to K^{\pm}X)\times{\mathcal B}(X\to\eta_c\pi^0)<5.7\times 10^{-6}$ for $X=X(3730)$ and $<1.2\times 10^{-5}$ for $X=X(4014)$.

We search for the neutral partners of the recently observed exotic states $Z(3900)^{\pm}$ and $Z(4020)^{\pm}$ and set 90\% C.L. upper limits on the product branching fractions ${\mathcal B}(B^{\pm}\to K^{\pm}Z)\times{\mathcal B}(Z\to\eta_c\pi^+\pi^-)$ of $4.7\times10^{-5}$ for $Z=Z(3900)^0$ and $1.6\times10^{-5}$ for $Z=Z(4020)^0$.

We set 90\% C.L. upper limits on the following $B$ decays involving the state $X(3915)$, whose origin is still unknown:  ${\mathcal B}(B^{\pm}\to K^{\pm}X(3915))\times{\mathcal B}(X(3915)\to\eta_c\eta)<3.3\times 10^{-5}$ and ${\mathcal B}(B^{\pm}\to K^{\pm}X(3915))\times{\mathcal B}(X(3915)\to\eta_c\pi^0)<1.8\times 10^{-6}$.

There are no theoretical predictions for the decay branching fractions of $D^0\bar{D}^{*0}$ molecular states similar to $X(3872)$. In this paper, we obtain an upper limit on the product branching fraction ${\mathcal B}(B^{\pm}\to K^{\pm}X_1(3872))\times{\mathcal B}(X_1(3872)\to\eta_c\pi^+\pi^-)$, which is of the same order as the product branching fraction  ${\mathcal B}(B^{\pm}\to K^{\pm}X(3872))\times{\mathcal B}(X(3872)\to J/\psi\pi^+\pi^-)$ measured in Ref.~\cite{concl1}. A similar situation is observed with the upper limit on the ${\mathcal B}(B^{\pm}\to K^{\pm}X(3915))\times{\mathcal B}(X(3915)\to\eta_c\omega)$ and the value of ${\mathcal B}(B^{\pm}\to K^{\pm}X(3915))\times{\mathcal B}(X(3915)\to J/\psi\omega)$ obtained in Ref.~\cite{x39154}. A more copious data set expected from the upcoming Belle II experiment~\cite{belleII} can provide an opportunity to determine the ratios of the decay branching fractions of the exotic states described above.

\section*{Acknowledgments}

We thank the KEKB group for the excellent operation of the
accelerator; the KEK cryogenics group for the efficient
operation of the solenoid; and the KEK computer group,
the National Institute of Informatics, and the 
PNNL/EMSL computing group for valuable computing
and SINET4 network support.  We acknowledge support from
the Ministry of Education, Culture, Sports, Science, and
Technology (MEXT) of Japan, the Japan Society for the 
Promotion of Science (JSPS), and the Tau-Lepton Physics 
Research Center of Nagoya University; 
the Australian Research Council and the Australian 
Department of Industry, Innovation, Science and Research;
Austrian Science Fund under Grant No.~P 22742-N16 and P 26794-N20;
the National Natural Science Foundation of China under Contracts 
No.~10575109, No.~10775142, No.~10875115, No.~11175187, and  No.~11475187; 
the Ministry of Education, Youth and Sports of the Czech
Republic under Contract No.~LG14034;
the Carl Zeiss Foundation, the Deutsche Forschungsgemeinschaft
and the VolkswagenStiftung;
the Department of Science and Technology of India; 
the Istituto Nazionale di Fisica Nucleare of Italy; 
National Research Foundation (NRF) of Korea Grants
No.~2011-0029457, No.~2012-0008143, No.~2012R1A1A2008330, 
No.~2013R1A1A3007772, No.~2014R1A2A2A01005286, No.~2014R1A2A2A01002734, 
No.~2014R1A1A2006456;
the Basic Research Lab program under NRF Grant No.~KRF-2011-0020333, 
No.~KRF-2011-0021196, Center for Korean J-PARC Users, No.~NRF-2013K1A3A7A06056592; 
the Brain Korea 21-Plus program and the Global Science Experimental Data 
Hub Center of the Korea Institute of Science and Technology Information;
the Polish Ministry of Science and Higher Education and 
the National Science Center;
the Ministry of Education and Science of the Russian
Federation and the Russian Federal Agency for Atomic Energy;
the Slovenian Research Agency;
the Basque Foundation for Science (IKERBASQUE) and 
the Euskal Herriko Unibertsitatea (UPV/EHU) under program UFI 11/55 (Spain);
the Swiss National Science Foundation; the National Science Council
and the Ministry of Education of Taiwan; and the U.S.\
Department of Energy and the National Science Foundation.
This work is supported by a Grant-in-Aid from MEXT for 
Science Research in a Priority Area (``New Development of 
Flavor Physics'') and from JSPS for Creative Scientific 
Research (``Evolution of Tau-lepton Physics'').

\end{document}